\documentclass[aip, reprint, author-year, floatfix]{revtex4-1}

\usepackage{gensymb}
\usepackage{graphicx}
\usepackage[caption=false]{subfig}
\usepackage[american]{babel}
\usepackage{siunitx}
\usepackage{amssymb}
\usepackage{booktabs}
\usepackage{dcolumn}
\usepackage{xcolor}
\usepackage[normalem]{ulem}
\usepackage{acronym}

\newacro{AC}[AC]{alternating current}
\newacro{DBD}[DBD]{dielectric-barrier-discharge}
\newacro{AC-DBD}[AC-DBD]{alternating-current dielectric-barrier-discharge}
\newacro{GTS}[GTS]{Ground Transportation System}
\newacro{PIV}[PIV]{particle image velocimetry}
\newacro{TKE}[TKE]{turbulent kinetic energy}
\newacro{TRL}[TRL]{Technology Readiness Level}
\newacro{LIC}[LIC]{line integral convolution}

\draft 

\begin{document}


\title{Drag reduction via separation control using plasma actuators on a truck cabin side} 



\author{Lucas Schneeberger}
    \email{lucas.schneeberger@uc3m.es}
\author{Stefano Discetti}
\author{Andrea Ianiro}
\affiliation{Department of Aerospace Engineering, Universidad Carlos III de Madrid, Leganés, Spain}


\date{\today}

\begin{abstract}
We investigate the drag reduction on a heavy-duty vehicle using dielectric-barrier discharge plasma actuators located on the A-pillars. An experimental campaign is carried out on a generalized truck model, the \ac{GTS}, which is known for its lateral separation bubbles on both sides of the truck's cabin. Measurements are performed for several yaw angles up to $7.5\degree$. Actuation is applied individually on the leeward and windward sides as well as simultaneously.
Load cell measurements show that the plasma actuators effectively reduce the axial force on the \ac{GTS}, with symmetric actuation achieving the highest reduction. Leeward actuation demonstrates greater control authority than the windward one; at large yaw angles the latter has a negligible effect on the axial force. Regarding side force, the leeward actuation produces a drop in its magnitude while windward actuation produces an increase. Interestingly, actuating symmetrically also augments the side force. Particle image velocimetry reveals that the plasma actuator causes a reduction in the length and width of the separation bubble on the cabin side, reducing the apparent frontal area of the truck and thus its drag. Under crosswind conditions, the stronger authority of the leeward actuator is explained by the larger separation bubble. The side force variation is driven by the net lateral suction force, which correlates with the size of the lateral recirculation regions controlled by the actuators.
\end{abstract}

\acresetall


\maketitle 

\section{Introduction}
\label{sec:introduction}

Since the 1970s oil crisis, fuel economy has emerged as a priority in transport: firstly because of economic concerns and more recently for environmental reasons. Nowadays, despite representing only 2\% of the European vehicles, trucks are responsible for 25\% of CO\textsubscript{2} emissions caused by road transport. The heavy-duty fleet even accounts for 6\% of global EU greenhouse gas emissions \citep{marcu_review_2023}.
As heavy vehicles mainly operate at speeds above 80 km/h, aerodynamics is a major source of fuel consumption. Consequently, reducing their aerodynamic drag is a key area of research.

The early work by \cite{saltzman_drag_1974} proved that dramatic drag reduction could be obtained by rounding the edges of a box-shaped van. \cite{hoerner_pressure_1965} had predicted that forebody drag reduction and base pressure were connected, so researchers also started to look into solutions for reducing drag through base pressure increase. \cite{peterson_drag_1981} obtained a 32\% drag reduction on a box-shaped van using an ogival rear fairing.
A decade later, researchers at Sandia Labs obtained an average drag reduction of 20\% using similar boat-tailing devices on a standardized truck model, the \ac{GTS} model, at several yaw angles. The \ac{GTS} was originally introduced by \cite{gutierrez_aerodynamics_1996} and was later exhaustively characterized at $Re=2 \times 10^6$ \citep{storms_experimental_2001}.

Cab deflectors, side skirts, and roof fairings progressively became widely adopted on long-haul as well as smaller delivery trucks. \cite{park_2024_2024}, in a study conducted on 26 major fleet operators covering 5\% of the heavy-duty vehicles in North America, showed that 90\% of the trucks now use trailer skirts. However, despite encouraging results, boat tail plates and other wake reduction devices are far from being frequent on heavy-duty vehicles.

One of the drawbacks of these passive devices is that they are designed for given nominal conditions. Overcoming this limitation, technological advances have enabled the development of active flow control systems that can adapt to external conditions. Jets are a common example of such actuators that have successfully been used for drag reduction on heavy-duty vehicles. \cite{englar_advanced_2001} pioneered their use in truck aerodynamics, achieving net power savings of up to 43\%. Furthermore, they investigated the capability of blowing devices to generate moments to act on directional stability. \cite{pfeiffer_robust_2018} and \cite{semaan_aerodynamic_2023} recently pursued these efforts and developed advanced control strategies to enhance the aerodynamic performance of a shortened \ac{GTS} while maintaining lateral directional stability in several wind directions.

Since the early 2000s, \ac{DBD} plasma actuators have emerged as a new type of actuator to control fluid flows. Despite their relatively low induced velocity, they can energize a boundary layer, have a very fast response time and can seamlessly be embedded into larger bodies, making them an ideal candidate for flow separation control. Several types of \ac{DBD} plasma actuators have successfully been used to control flow separation. \cite{benard_benefits_2011} used a single linear \ac{DBD} fed by \ac{AC} to control separation over an airfoil. 
\cite{roy_dielectric_2016} and \cite{wang_aerodynamic_2022} placed different geometries of actuators on the top and the sides of a trailer, and respectively achieved 15\% and 9\% drag reduction with comb-shaped actuators, arguably capable of acting on a wider range of separation locations than a simple linear actuator.
Another interesting type of \ac{DBD} are the \unit{\nano \second}- and \unit{\micro \second}-pulsed plasma actuators. They provide higher control authority for separation control due to intense transient energy deposition and shock-induced momentum transfer and have recently reached enough maturity for integration in real flight applications \citep{su_uav_2018}. However, their very fast voltage rise may produce interferences in electronic devices and induce current in nearby conducting materials; furthermore, the high-voltage signal can reflect on the discontinuities of the transmission cables \citep{kotsonis_diagnostics_2015}. Consequently, they require careful design of the electric supply circuit and prevention of electromagnetic interferences.

 Early studies from the 1980s \citep{tyrrell_aerodynamics_1987, drollinger_heavy_1987} claim that 40\% to 50\% of the drag is generated at the tractor front. \cite{vernet_separation_2015} also estimate the tractor contribution to be 50\% of the total drag at zero yaw angle. \cite{bonnavion_use_2022} showed that turbulent production in the frontal regions of bluff bodies diminishes the pressure recovery at the base and therefore increases drag. The flow separation on the front corner of the trailer (the so-called A-pillar) was already targeted in the 1980s when truck manufacturers started rounding the sharp front corners. We consider this part of the drag especially easy to target because tractor modifications are independent of the type of payload and may be implemented directly by the tractor manufacturers. Also, as pointed out by \cite{cooper_truck_2003}, tractor-mounted devices have a shorter payback time as there are several trailers for a given tractor.

Even though many truck models exhibit a lateral separation bubble that can be controlled, the \ac{GTS} appears as the most generalizable one to extend any new approach to other flow control cases. It is also the most commonly studied, which allows for results comparison. \cite{mcarthur_near_2016} also studied the \ac{GTS} at a lower Reynolds number ($Re=2.7 \times 10^4$), aiming to provide an experimental baseline for assessment of numerical simulations. They observed similar flow features as \cite{storms_experimental_2001}, in particular the presence of lateral separation bubbles for several yaw angles. \cite{rao_investigation_2019} also showed the presence of a separation bubble on both sides of the \ac{GTS} cabin with numerical simulations at that same $Re$.

\cite{minelli_numerical_2016} conducted a numerical investigation of the flow past the A-pillar of a generalized truck and showed that zero-net mass-flux synthetic jets can be employed to reduce the size of the separation region and consequently the drag. They observed significant performance changes depending on the actuation frequency. They confirmed these findings in an experimental follow-up study \citep{minelli_aerodynamic_2017}.

Crosswind affects the size of the separation bubble, increasing it on the leeward side and reducing it on the windward side \citep{rao_investigation_2019}.
Targeting the larger separation bubble, \cite{vernet_flow_2018} applied comb-shaped plasma actuators on the leeward A-pillar of a commercial truck to prevent separation. They obtained an encouraging 20\% of drag reduction at $9\degree$ yaw.
However, according to \cite{rao_investigation_2019}, flow separation is still present on the windward side at incidence angles up to $\alpha=10 \degree$. This motivates considering both the leeward and windward sides even in crosswind conditions, although differentiating them in the applied control strategy.

We propose using linear \ac{AC-DBD} plasma actuators (hereafter simply referred to as \ac{DBD}) to control the lateral separation bubble on the A-pillar of a \ac{GTS} truck model. Early results from \cite{vernet_separation_2015} support the hypothesis that a linear \ac{DBD}, i.e. a straight electrode on each side of the dielectric material, may be sufficient to control the separation past the A-pillar while preserving manufacturing and operational simplicity. From the same study, it appears that the linear actuator does not necessarily have to be positioned at the exact location of the separation, as it is still remaining effective over a significant range around the separation point. Their later study \citep{vernet_flow_2018} used comb-shaped vortex generators to mitigate the separation over the A-pillars of a truck. However, this geometry induces a wall-normal flow, which in turn confers stronger three-dimensional dynamics to the bubble, including the reattachment location. The flow topology is then significantly more difficult to study. In addition, they only considered leeward actuation. As separation also occurs on the windward side of the \ac{GTS}, we investigate the impact of placing actuators on both A-pillars.

Considering the above, the following research questions emerge:
\begin{itemize}
    \item Can the lateral separation bubble located after the A-pillars of the \ac{GTS} be controlled with linear \ac{DBD} plasma actuators?
    \item How does this actuation impact the forces at different yaw angles?
    \item Is asymmetric actuation favorable in yawed conditions?
\end{itemize}

To answer them, we performed wind tunnel experiments on a \ac{GTS} model equipped with plasma actuators on both A-pillars, measuring the axial and lateral forces in the non-actuated and actuated cases. In section \ref{sec:methodology} we describe the experimental setup and the measurement technique employed. In section \ref{sec:forceresults} we present the results from the force measurement campaigns. In an attempt to explain the force variations with changes in the flow fields, we report the velocity fields obtained with \ac{PIV} experiments in section \ref{sec:PIVresults}. The impact of the various actuation strategies in different wind directions is discussed, and the article is concluded with the suggestion of an actuation policy.

\section{Methodology}
\label{sec:methodology}

\begin{figure*}
    \centering
    \begin{minipage}[t]{12 cm}
        \subfloat[]{\includegraphics[width=\linewidth]{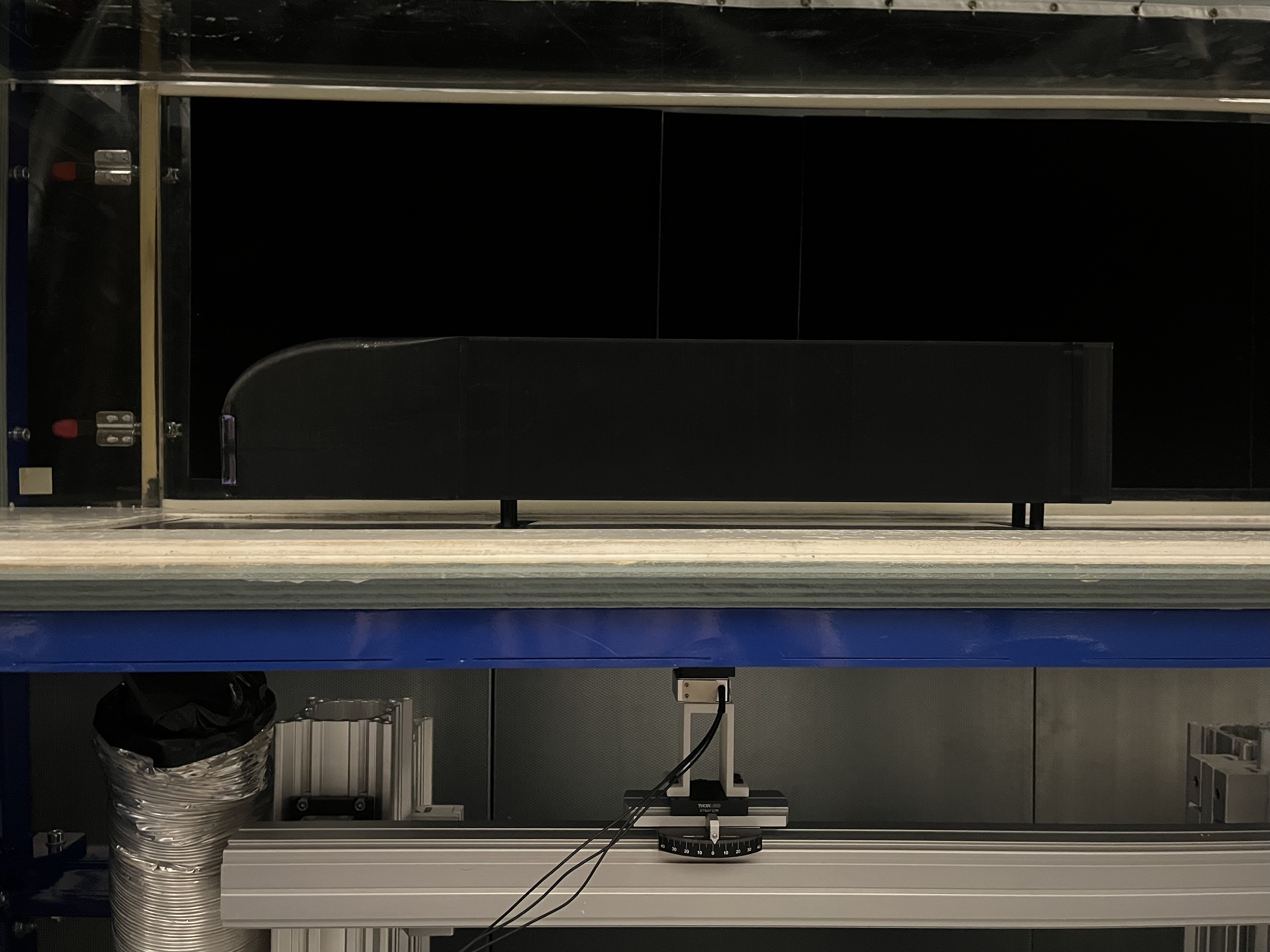}
        \label{fig:setup_picture}}
    \end{minipage}%
    \hfill
    \begin{minipage}[t]{5cm}
        \subfloat[]{
        \includegraphics[width=\linewidth]{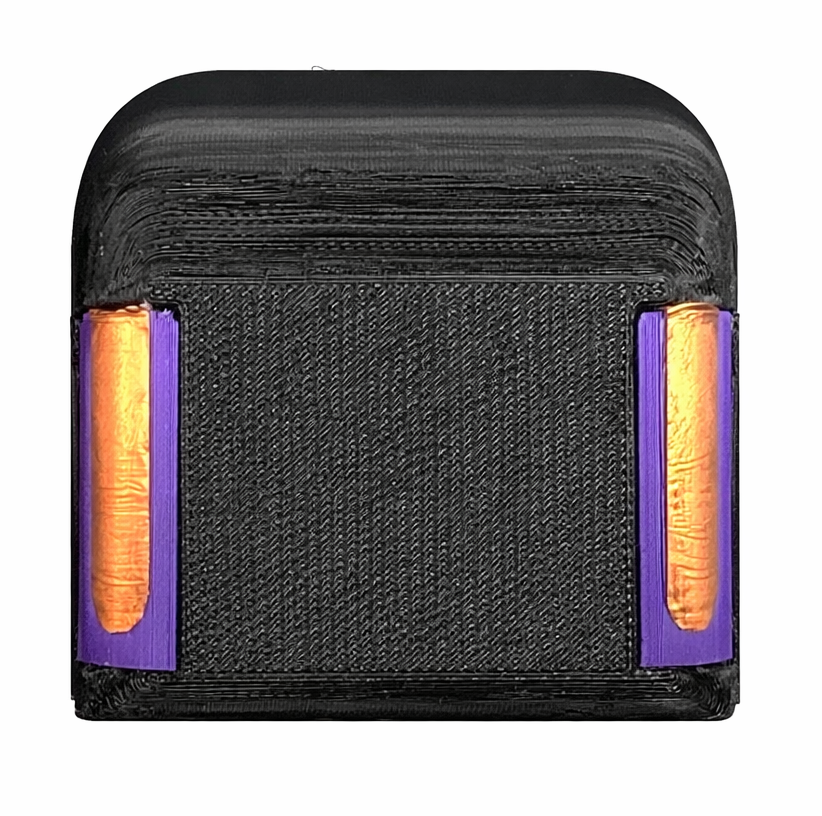}
        \label{fig:PA_zoom}}
    \end{minipage}
    \caption{(a): Picture of the GTS model placed in the wind tunnel. The boundary layer suction duct and the load cell appear below the wind tunnel floor. (b): Detail of the DBD plasma actuators embedded on the truck's A-pillars.}
    \label{fig:setup_pictures}
\end{figure*}

\begin{figure*}[h]
    \centering
    \includegraphics[width=\linewidth]{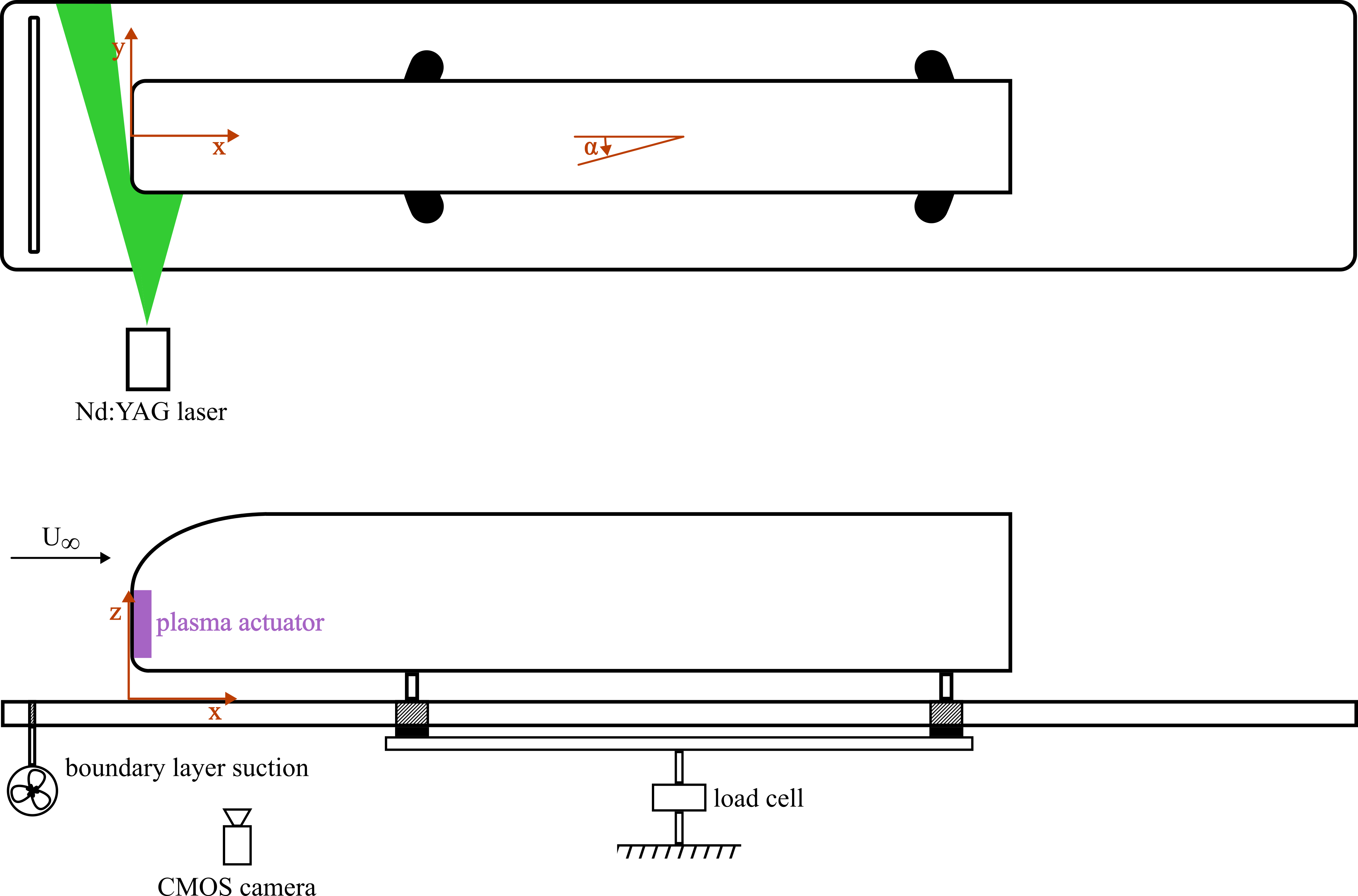}
    \caption{Sketch of the experimental setup. The reference axes are in red and the laser sheet is in green. The plasma actuator indicated in violet is also present on the other side.}
    \label{fig:setup_sketch}
\end{figure*}

\begin{figure}
    \centering
    \includegraphics[width=0.8\linewidth]{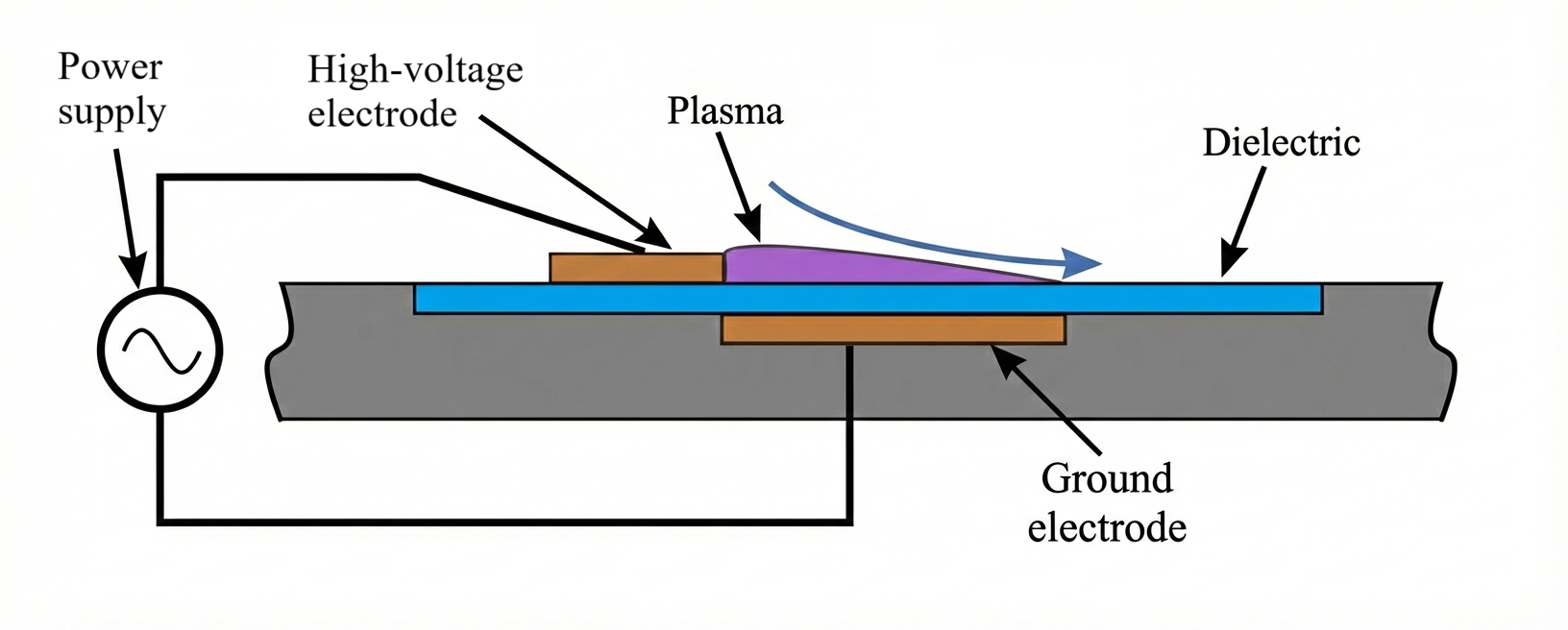}
    \caption{Functioning principle of a \ac{DBD} plasma actuator.}
    \label{fig:sketch PA}
\end{figure}

The experiments are conducted in the closed-loop wind tunnel of the Department of Aerospace Engineering of the University Carlos III of Madrid. The tunnel has a test section of $40~\unit{\centi \meter}$ x $40~\unit{\centi \meter}$ and a maximum wind velocity of $20~\unit{m/s}$. The turbulence intensity is below 1\%. A picture of the model inside the wind tunnel is shown in Figure \ref{fig:setup_picture}.
The \ac{GTS} size is adapted to the dimensions of the wind tunnel cross-section, to avoid exceeding 15\% blockage ratio at $10\degree$ yaw. This yields a model width of $W=8.5~\unit{\centi \meter}$, height of $H=11.8~\unit{\centi \meter}$, and length of $L=65~\unit{\centi \meter}$.

The wind tunnel floor consists of a methacrylate plate with holes for the truck supports and a slit for boundary layer suction of width $6~\unit{\milli \meter}$ ($0.07~\unit{W}$), whose centerline is at $x/W=-0.85$ with respect to the origin, as shown in Figure \ref{fig:setup_sketch}.
The truck yaw is modified via curved supporting modules visible in black in the upper view of the sketch. They are pierced with holes for the truck's cylindrical legs at nominal angles of $0\degree$, $2.5\degree$, $5\degree$, $7.5\degree$ and $10\degree$. As shown by \cite{lawson_effects_2008}, the relative yaw angle in highway conditions is practically contained between $-10\degree$ and $10\degree$. However, only angles up to $\pm7.5\degree$ can be tested, as for $\pm10\degree$, the side force induces truck tilting. The alignment precision is about $0.3\degree$ and, to ensure repeatability, a yaw angle correction is performed based on force measurements, as will be detailed hereinafter.

Experiments are performed at a Reynolds number based on the truck width of $Re=4.2\times 10^4$. This value corresponds to a wind-tunnel speed of approximately $U_\infty=8.3~\unit{m/s}$, which has been selected based on the expected control authority of the plasma actuators.
The dynamic pressure $q=\frac{1}{2}\rho U_\infty^2$ is measured before each experiment by placing a spanwise-centered Pitot tube upstream of the methacrylate plate depicted in Figure \ref{fig:setup_sketch}. The pressure is measured using a SurreySensors pressure scanner with a range of $160~\unit{Pa}$. The corresponding uncertainty is $0.4~\unit{Pa}$.
Since the wind tunnel is operated at constant power, the dynamic pressure is corrected using the temperature evolution during the force experiments. For the Reynolds number estimation, air viscosity is evaluated using Sutherland's formula. The temperature is measured using a thermocouple placed inside the tunnel downstream of the model, with an uncertainty of less than $1~\unit{K}$.

The following subsections present the arrangement of the plasma actuators (section \ref{sec:plasma}), the details of the force measurements (section \ref{sec:forces}), and the procedure for \ac{PIV} measurements and analysis (section \ref{sec:piv}).

\subsection{Plasma actuation}
\label{sec:plasma}

The functioning principle of \ac{DBD} plasma actuators is sketched in Figure \ref{fig:sketch PA}. A plasma jet is generated due to a voltage difference between two electrodes separated by a dielectric material. We use 3D-printed PLA quarter-cylinders of $1~\unit{\milli \meter}$ thickness as the dielectric and apply a 10-mm-wide copper tape on both sides to act as electrodes. The edges of the electrodes are cut round by hand to avoid electric field concentration and arc discharges. The electrodes are offset in the flow direction so that they overlap by $5~\unit{\milli \meter}$. Plasma is generated at the trailing edge of the outer electrode, corresponding to an azimuthal location on the quarter-cylinder of approximately $45\degree$ from the truck's main axis. As shown on Figure \ref{fig:PA_zoom}, the dielectric and lower electrode of the plasma actuator are embedded in each A-pillar of the \ac{GTS}, both of which have been previously trimmed so that the dielectric is flush-mounted with the rest of the truck cabin.

We supply the two actuators with a $13~\unit{kHz}$ AC signal generated by two GBS Elektronik Minipuls 4. Higher actuation frequencies have been tested, but they lead to mutual inductance between the two actuators, making it difficult to generate actuation on only one side confidently. The cables delivering the current to the actuators are routed inside the model's body, so no to influence the aerodynamics. The two Minipuls transformers are fed with the same direct voltage ($13.5~\unit{V}$). However, their output signal, which is then fed to the actuators, is slightly different: $8.7~\unit{kVpp}$ on average for the left actuator and $10.4~\unit{kVpp}$ on average for the right one. The difference might be due to internal losses inside the electrical circuit of the right actuator, as the measured side force when the truck is at $0\degree$ yaw and both actuators are turned on is within uncertainty limits.

The velocity induced by the plasma actuator is estimated to be close to $4~\unit{m/s}$. A close-up \ac{PIV} experiment has been carried out at a yaw angle of $\alpha=5\degree$ with and without actuation. The difference between the actuated and non-actuated velocity fields $\Delta U$ has been computed, and a region very close to the actuator stands out for its added velocity when actuation is on. The maximum of this region provides an estimate of the plasma-induced velocity consistent with the results of \cite{forte_optimization_2007}.

Each actuator consumes about $1.3~\unit{W}$. This value has been measured with the shunt capacitor technique \citep{kotsonis_diagnostics_2015}. This refers to the electrical power consumed and not the mechanical work injected in the flow, which includes some parasitic losses. We find this value to be reasonably approximated by the model proposed by \cite{wilde_model_2021}.


\subsection{Force measurements}
\label{sec:forces}

Aerodynamic forces are measured with a Fibos FA702 3-axis load cell placed below the test section floor and attached to the truck model through its four legs, as sketched in Figure \ref{fig:setup_sketch}. The load cell signal is amplified and digitized by three Fibos FA11 transmitters, one per axis. The force signal is then acquired at a frequency between $20$ and $30~\unit{Hz}$.

Before the experiments, the load cell gain has been calibrated individually for each axis, preventing cross-talk between axes. For the load cell zeroing, two measurements are acquired with the wind tunnel off, both before and after each experiment. The two wind-off measurements (resulting from averaging timeseries) are averaged to minimize the effects of any thermal drift in the load cell during the force measurement campaign.
Each force measurement (with the wind tunnel on) is repeated 3 times to minimize uncertainty. Each measurement is obtained as the average of 1000 samples.

For non-zero yaw angles, we perform axial and lateral force measurements under four actuation conditions: actuators off, both actuators functioning simultaneously (\textit{symmetric actuation}), only leeward actuation, and only windward actuation. For $0\degree$ yaw, the left and right actuators have an equivalent effect, so leeward and windward actuations are simply called \textit{asymmetric actuation} and force measurements are performed only for non-actuated, symmetric, and asymmetric conditions.

The load cell is aligned with the flow direction, so the forces are measured in the wind reference frame and are projected to obtain the forces in the truck's reference frame, which is of relevance for road applications.
The force coefficients are computed as $C_x = F_x/qA$ and $C_y = F_y/qA$ where $A$ is the \ac{GTS} frontal area.  

The uncertainty of force measurements can be estimated following \cite{moffat_describing_1988} employing the uncertainties of intermediate variables reported in Table \ref{tab:uncertainties}. This approach is consistent with the general framework recommended by the International Standardization Organization \citep{jcgm_evaluation_2008}.

\begin{table}
\caption{Uncertainties on the force coefficients and the intermediary variables.}
\label{tab:uncertainties}
\begin{tabular}{cl}
 variable & \text{uncertainty}\\
\hline
$T$ & $1~\unit{K}$ \\
$\alpha$ & $0.29\degree$ \\
$q$ & $0.4~\unit{Pa}$ \\
$Re$ & $410$ \\
$D, S$ & $0.010~\unit{N}$ \\
$F_x$ & $0.011~\unit{N}$ \\
$F_y$ & $0.010~\unit{N}$ \\
$C_x$ & $0.031$ \\
$C_y$ & $0.029$ \\
$\Delta C_x$ & $0.0088$ \\
$\Delta C_y$ & $0.0089$ \\
\hline
\end{tabular}
\end{table}

 %

 The force coefficient variation due to plasma actuation is computed for each axis $i$ as $\Delta C_i = |C_i^{ACT}| - |C_i^{NA}|$. The $ACT$ and $NA$ superscripts indicate actuated and non-actuated conditions, respectively.
 For the uncertainty quantification of this quantity, since we are interested in the variation of the actuated coefficient relative to the non-actuated one, we can discard the biases that affect both variables in the same way, e.g. thermal effects in the load cell. This results in a relative uncertainty lower than that of the absolute value of the coefficients, as can be observed in the last two rows of Table \ref{tab:uncertainties}.

We validate our testing procedure by applying it to an experiment at $Re=2.7 \times 10^4$ to compare with \cite{rao_investigation_2019}. In this test, we only consider the baseline axial force coefficients $C_x^{NA}$ for yaw $0\degree$ and $2.5\degree$. After correcting for blockage effects, we obtain $C_x=0.58$ at $0\degree$ and $C_x=0.60$ at $2.5\degree$, which is consistent with Rao et al.'s results ($0.56$ and $0.58$, respectively). The applied blockage correction follows the methodology proposed by \cite{ewald_bluff-body_1998}, i.e. $C_{x,c} = \frac{C_{x,u}}{1+2.41 C_{x,u}(S/C)}$, where $u$ and $c$ subscripts indicate uncorrected and corrected values and $S/C$ denotes the blockage ratio. In the remainder of this paper, since the main purpose is to compare non-actuated and actuated forces under the same blockage conditions, blockage corrections are not applied.

Given the uncertainty in model alignment and along the measurement chain, we compute a linear regression of the non-actuated $C_x(\alpha)$ and $C_y(\alpha)$ of all the collected runs. A linear dependence of the force coefficients on yaw angle is expected, in agreement with previous studies \citep{storms_experimental_2001, rao_investigation_2019}, given that the investigated angles are small ($\alpha < 10^\circ$). A RMSE minimization allows a correction for the yaw angle $\alpha$ based on the deviation of the measured coefficient from the regression model. The experiments whose deviation from the fit is larger than the measurement uncertainty are discarded, starting with those with greater deviation. The regression is recomputed after every discard, leading to an iterative process until all remaining experiments have a distance to the regression smaller than their uncertainty.

Once left with this remaining set of experiments, the effect of the actuation is considered, and runs whose values of $\Delta C_x$ or $\Delta C_y$ are outliers within their own series of 3 runs are discarded. 
As a final check, linear regressions for each actuation mode (symmetric, leeward, and windward) are computed, ensuring that no data point deviates from the regression by more than its uncertainty. A total of $83$ runs is presented and discussed in the following.

\subsection{PIV}
\label{sec:piv}

\ac{PIV} experiments are conducted in the horizontal plane at $z/W=0.65$ above the tunnel floor to gain further insight into the bubble topology and its dependence on actuation. The horizontal measurement plane is sketched in the top view of Figure \ref{fig:setup_sketch}. The plane height is chosen considering the separation region characterized by \cite{rao_investigation_2019}, who describes a convex-shaped bubble, with a peak extension around mid-height. To confirm this topology in our case, we conducted an experiment with tufts on the cabin side at several yaw angles and adjusted the laser sheet height to capture a plane where the bubble reaches an extension close to its maximum for all yaws.
In the streamwise axis, the observation region starts $2~\unit{\milli \meter}$ upstream of the truck nose ($x/W$=-0.024) and extends over $135~\unit{\milli \meter}$ downstream ($x/W$=1.588). In the spanwise axis, the \ac{PIV} plane edge lies on the truck cabin and extends over 36 mm ($y/W$=0.424). These dimensions have been chosen based on the expected bubble size given the results of \cite{storms_experimental_2001}.

Di-ethylhexyl sebacate particles of approximately $1~\unit{\micro \meter}$ diameter and generated with a Laskin nozzle are employed to seed the flow. A double-pulsed Nd:YAG laser is used to illuminate the particles. Each of the two cavities of the laser produces a maximum pulse energy of $200\unit{\milli\joule}$. The delay between the two pulses $\Delta t$ is adjusted to obtain a displacement based on the incoming velocity $U_\infty$ below 10 pixels. The frequency at which the same cavity is triggered again is set to $f=14~\unit{Hz}$. An Andor SCMOS camera with a 2560 x 2160 pixel sensor ($6.5~\unit{\micro \meter}$ pixel size) is placed below the methacrylate tunnel floor. It is equipped with a Nikon objective of focal length $f= 50~\unit{\milli \meter}$. The resulting spatial resolution is approximately $16~\unit{px/mm}$. A total of 1500 snapshot pairs are taken for each yaw angle setting.

Before \ac{PIV} analysis, we apply two preprocessing steps to the raw images. We remove background reflections using the eigenbackground algorithm from \cite{mendez_pod-based_2017}. We subtract the first 20 modes resulting from the proper orthogonal decomposition (POD) of the raw images. We then apply a mask of static particles (i.e., fake particles at the same location in all images) uniformly distributed within the area covered by the truck model. This results in setting the velocity field to 0 in this area, preventing the generation of incorrect vectors due to spurious correlations caused by the lack of particles where the truck body lies.

The particle images are processed with the open software PaIRS, developed at the University of Naples "Federico II" by \cite{astarita_pairs_2022}. For the final iteration of the convolution algorithm, we use an interrogation window size of 32 pixels with 75\% overlap. This yields at least 170 vectors per model width for all the experiments.
The processing software PaIRS relies for the vector validation on the universal outlier detection algorithm by \cite{westerweel_universal_2005} with a semi-kernel of 2, a threshold of 2, and an acceptable fluctuation level due to cross-correlation of $\varepsilon=0.5$.
In addition to those, we consider as outlier any vector $\mathbf{u}(x, y)$ that has any of its two components larger than $m+2\sigma+0.3~\unit{pixels}$ in absolute value, with $m$ and $\sigma$ being respectively the median and the standard deviation of the 1500 velocity vectors at this location. The constant 0.3 is added to account for the uncertainty of the results. This correction is particularly needed in regions with low turbulence intensity. For all angles, this process retains a minimum of 700 valid vectors (47\%), with numbers this low only occurring in the close vicinity of the wall. The rest of the field of view generally displays at least 1350 valid vectors (90\%).
To replace outliers, we apply a gappy POD reconstruction on the invalid locations, similar to the first algorithm described in \cite{gunes_gappy_2006}, using 100 modes for the reconstruction and retaining the invalid vector values as an initial guess. The algorithm is iterated only once.

\section{Results \& Discussion}

In this section, first, the aerodynamic force results of the wind tunnel tests are presented and discussed; the velocity fields resulting from the \ac{PIV} experiments are then analyzed, aiming to explain the observed force variations in terms of plasma effects on the separation bubble.

\subsection{Force measurements}
\label{sec:forceresults}

\begin{figure}[t]
    \centering
        \subfloat[]{\includegraphics{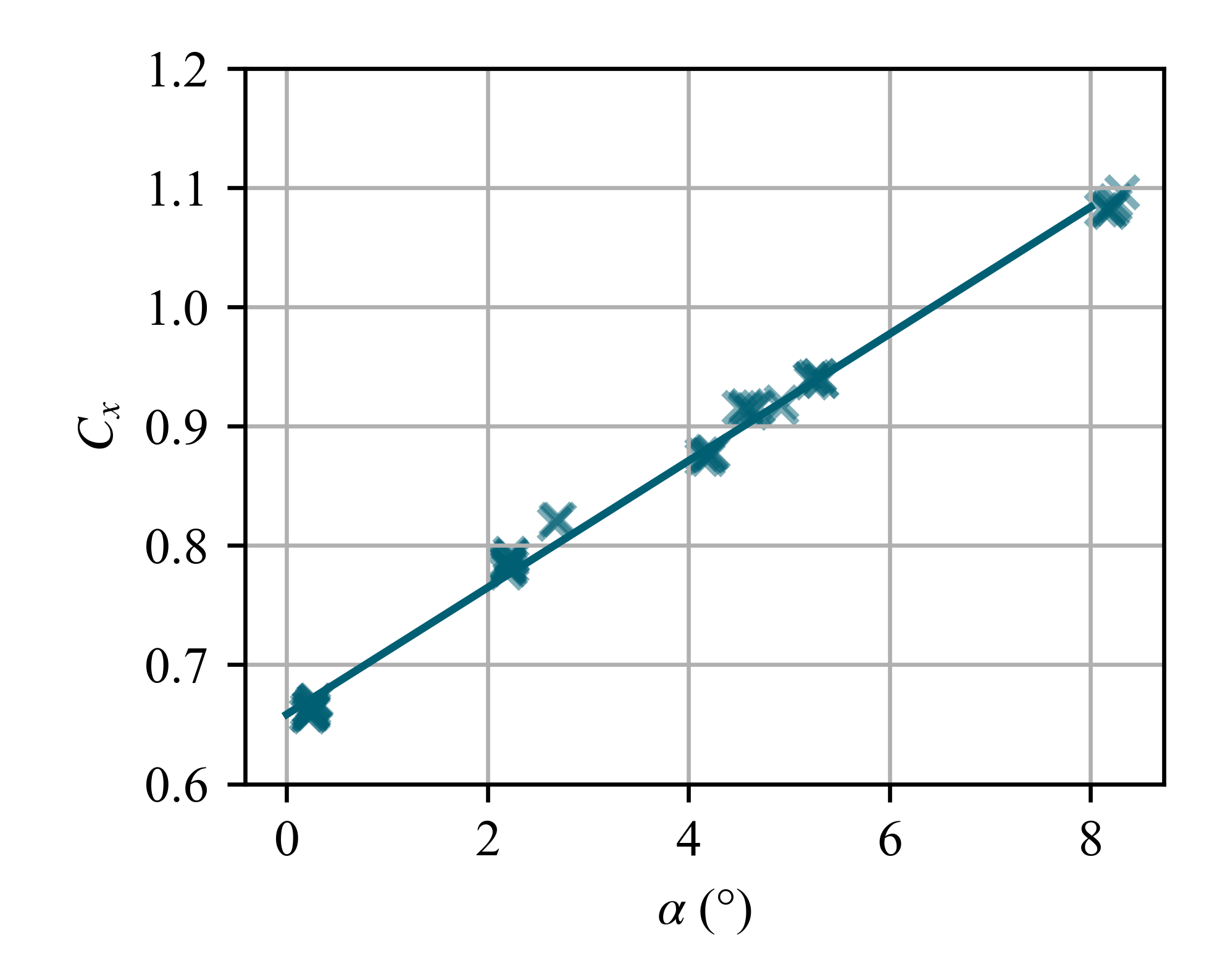}\label{fig:Cx_NA_alphacorr}}
        \hfill
        \subfloat[]{\includegraphics{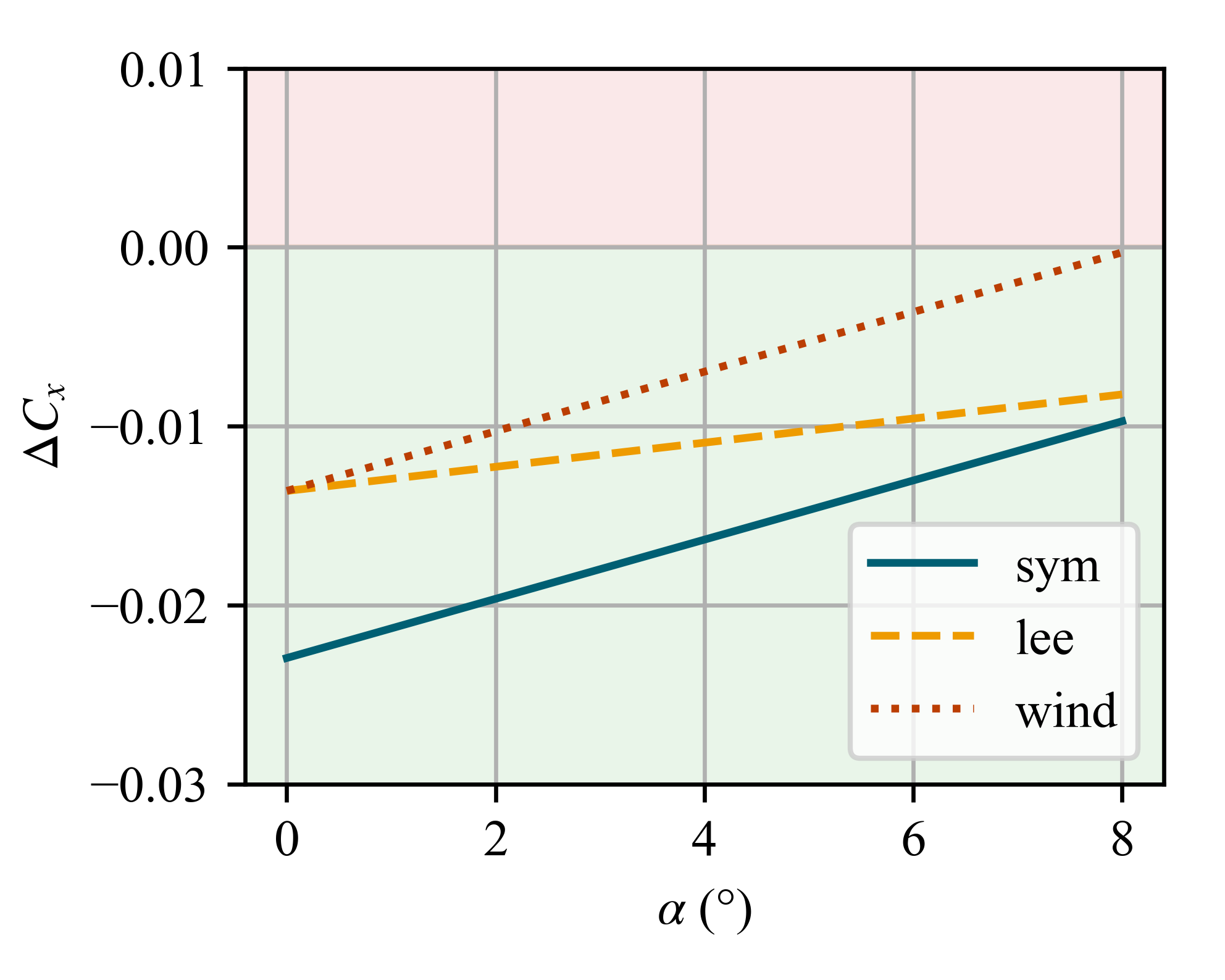}\label{fig:Delta_Cx_alphacorr}}
    \caption{Influence of the yaw angle on the axial force coefficient $C_x$. The experimental datapoints without actuation as well as the regression are shown in Figure \ref{fig:Cx_NA_alphacorr}. Figure \ref{fig:Delta_Cx_alphacorr} shows only the difference between the linear fit of the baseline case and the ones of the actuated cases. The area where drag is reduced is marked in green, while red indicates a drag increase.}
    \label{fig:Cx_alphacorr}
\end{figure}

\begin{figure}[t]
    \centering
        \subfloat[]{\includegraphics{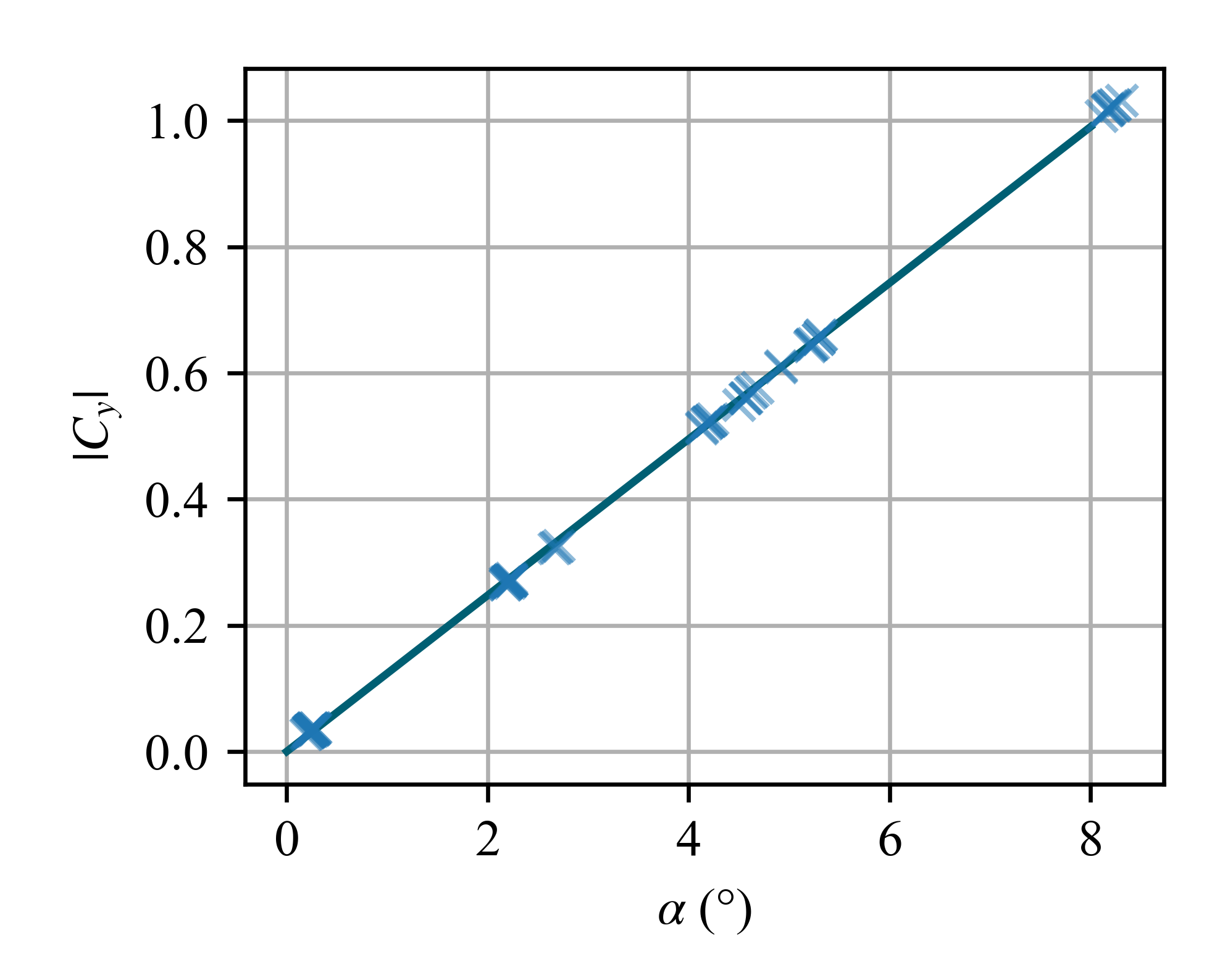}\label{fig:Cy_NA_alphacorr}}
        \hfill
        \subfloat[]{\includegraphics{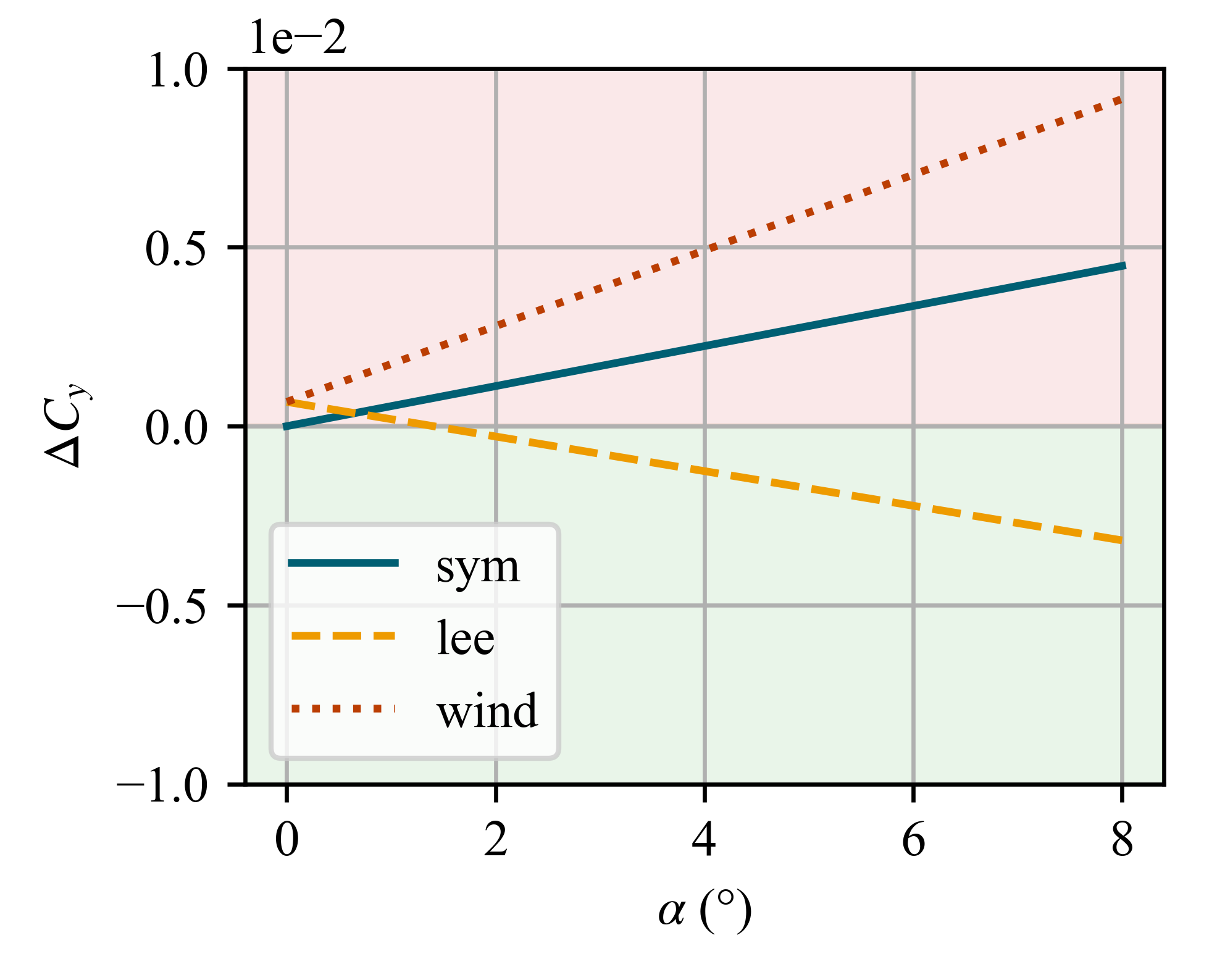}\label{fig:Delta_Cy_alphacorr}}
    \caption{Influence of the yaw angle on the absolute lateral force coefficient $|C_y|$. The experimental datapoints without actuation as well as the regression are shown in Figure \ref{fig:Cy_NA_alphacorr}. Figure \ref{fig:Delta_Cy_alphacorr} shows only the difference between the linear fit of the baseline case and the ones of the actuated cases. The area where the absolute lateral force is reduced is marked in green, while the area where it increases appears in red.}
    \label{fig:Cy_alphacorr}
\end{figure}


Figure \ref{fig:Cx_alphacorr} reports the effect of the yaw angle on the axial force coefficient for different actuation modes. In all cases, $C_x$ increases with the yaw angle. Figure \ref{fig:Cx_NA_alphacorr} presents the variation of the axial force coefficient of the non-actuated \ac{GTS} with $\alpha$. The axial force coefficient increases linearly with the yaw angle, from a value slightly larger than 0.65 at $0\degree$ to almost 1.1 at $8\degree$. This linear increase is consistent with previous literature results, such as the work by \cite{rao_investigation_2019} at $Re=2.7 \times 10^4$, and \cite{storms_experimental_2001} at $Re=2\times 10^6$.

Figure \ref{fig:Delta_Cx_alphacorr} reports the difference $\Delta C_x = |C_{x}^{ACT}| - |C_{x}^{NA}|$ between the fit of the data in Figure \ref{fig:Cx_NA_alphacorr} and the fits of the $C_x$ values measured for the actuated cases. The regressions are constrained so that the windward and the leeward actuation have the same offset at the origin, as the experimental data at $0\degree$ is the same. The results suggest that the actuators act practically independently: when actuated symmetrically, the resulting drag reduction is close to the sum of the leeward and windward effects. This is not surprising, since the actuators control the two separation bubbles, located on either side of the truck cabin, and thus are unlikely to influence each other.

Interestingly, the leeward actuation shows little sensitivity to the yaw angle regarding its drag reduction capability. On the other hand, the drag reduction achieved through windward actuation decreases sharply with increasing yaw angle, eventually leading to no effect on $C_x$ for $\alpha \gtrapprox 8 \degree$. When both actuations are combined, the drag reduction remains effective at all considered yaw angles, but the effect becomes smaller with increasing yaw angle with a slope similar to the windward-actuated $\Delta C_x$ ($1.67 \times 10^{-3} \deg^{-1}$ for windward vs. $1.65 \times 10^{-3} \deg^{-1}$ for symmetric). This is opposite to the results of \cite{vernet_plasma_2018}, who found an increased drag reduction with yaw angle. However, they used plasma actuators with a different geometry, sometimes called \textit{serrated} or \textit{vortex generators}, which are less sensitive to the separation location. This might explain the difference with the present results.

Until $\alpha \approx 8 \degree$, the symmetric actuation is best at reducing drag. It is important to note that the linear data regressions reported herein are valid only for the relatively small yaw angles studied in our experiments. According to the findings reported by \cite{storms_experimental_2001}, starting from $\alpha\approx8\degree$, the scaling of $C_x$ with the yaw angle is not linear anymore, and at larger angles $C_x$ should eventually decrease, as we are considering the force along the vehicle's body axis. It is therefore difficult to predict how the trend found until $\alpha=8 \degree$, where the effect of leeward and symmetric actuations comes very close together, would extend at larger yaws.

Figure \ref{fig:Cy_alphacorr} depicts the evolution with the yaw angle of the magnitude of the lateral force coefficient expressed in the vehicle reference frame. As for the axial force, one can see a linear increase of the non-actuated force with the yaw angle (Figure \ref{fig:Cy_NA_alphacorr}), going from $C_y=0$ at $0\degree$ up to $C_y\approx 1$ at $8\degree$. The orientation of the force is always towards the leeward side, and is symmetric for the negative and positive yaw angles.

On the bottom panel (Figure \ref{fig:Delta_Cy_alphacorr}), the variation of the absolute $C_y$ with respect to the baseline is represented. Here as well, the regressions for leeward and windward actuation are constrained to have the same offset at the origin, since they employ the same data points. At $0\degree$ yaw angle, asymmetric actuation results in a positive $\Delta C_y$, as it creates an asymmetric flow field, leading to a non-zero side force.
At non-zero yaw angles (beyond $\alpha \approx 1 \degree$), actuating on the leeward A-pillar appears as the only possible actuation able to reduce the lateral force. In general, actuating windward reinforces the lateral force, and this effect becomes more pronounced at larger yaw angles. When both sides are combined, the effect is still an increase in the lateral force at non-zero yaw angles: the windward actuation dominates. This effect grows linearly with the yaw angle, even though not as sharply as with the windward actuation only ($1.1 \times 10^{-3} \deg^{-1}$ for windward vs. $5.6 \times 10^{-4} \deg^{-1}$ for symmetric). The magnitude of the considered changes is small, since the recirculation region downstream of the A-pillar represents only a fraction of the total lateral area. However, the results are informative of the flow physics and indicate a trend that could be reinforced with improved actuator design or control parameters.

Considering the results of both axial and lateral force coefficients reported in this study, we observe that the windward actuation offers little benefit for either drag reduction or lateral stability, except when combined with the leeward one for yaw angles close to zero. An interesting actuation strategy could be:

\begin{itemize}
    \item actuating with both actuators when the vehicle is approximately straight to benefit from the highest drag reduction;
    \item defining a $\alpha_{cut}>1\degree$ , from which the windward actuator is turned off to reduce the lateral force while keeping a good drag reduction.
\end{itemize}

The exact value of $\alpha_{cut}$ is to be defined according to the weighting of the drag reduction objectives with respect to the lateral stability ones. In addition, the energy to generate the actuation is also saved by turning one actuator off, thus increasing the net power savings. The prevalence of this third objective can also be taken into account in the definition of $\alpha_{cut}$.

\subsection{PIV results}
\label{sec:PIVresults}

\begin{figure*}[t]
    \centering
    \includegraphics{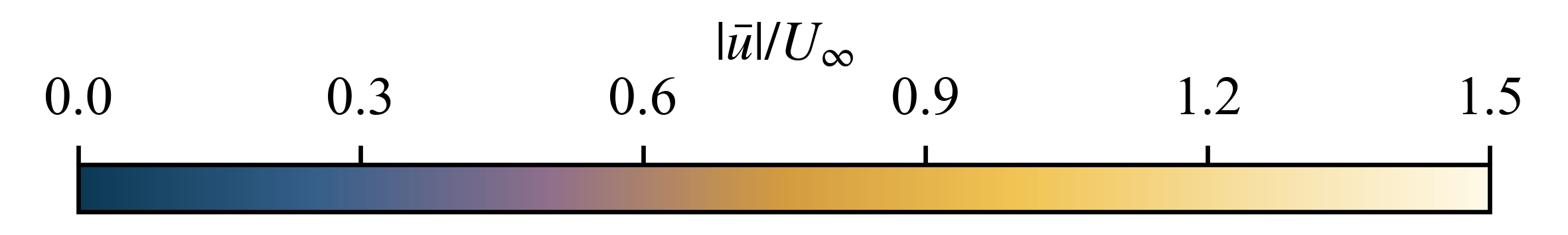}
\begin{minipage}{15.9cm}
    \centering
    \subfloat[]{\includegraphics[trim=0cm 0.7cm 0.1cm 0.1cm, clip]{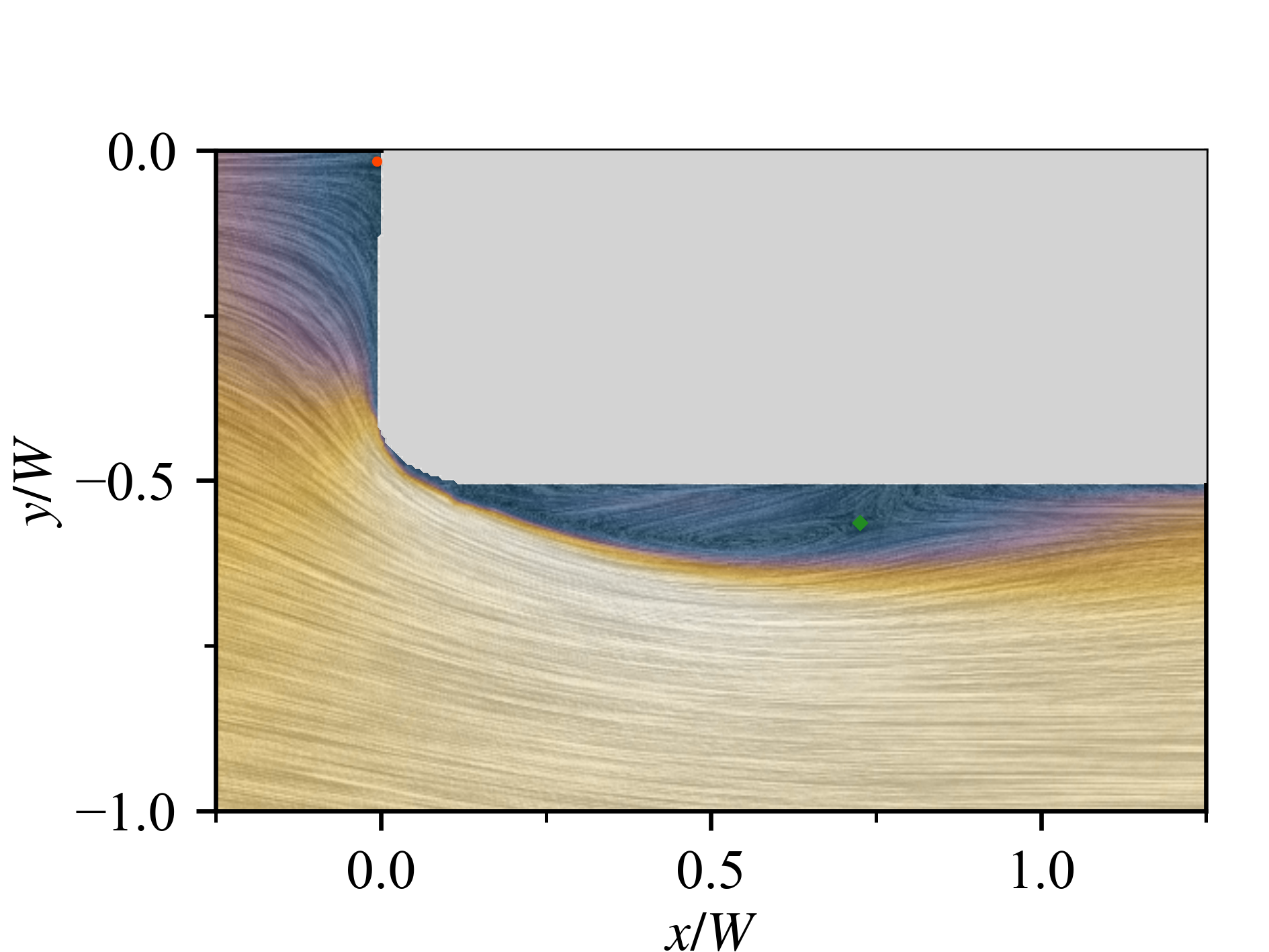}\label{fig:lic_avg_0deg_NA}}\hfill%
    \subfloat[]{\includegraphics[trim=1.3cm 0.7cm 0.1cm 0.1cm, clip]{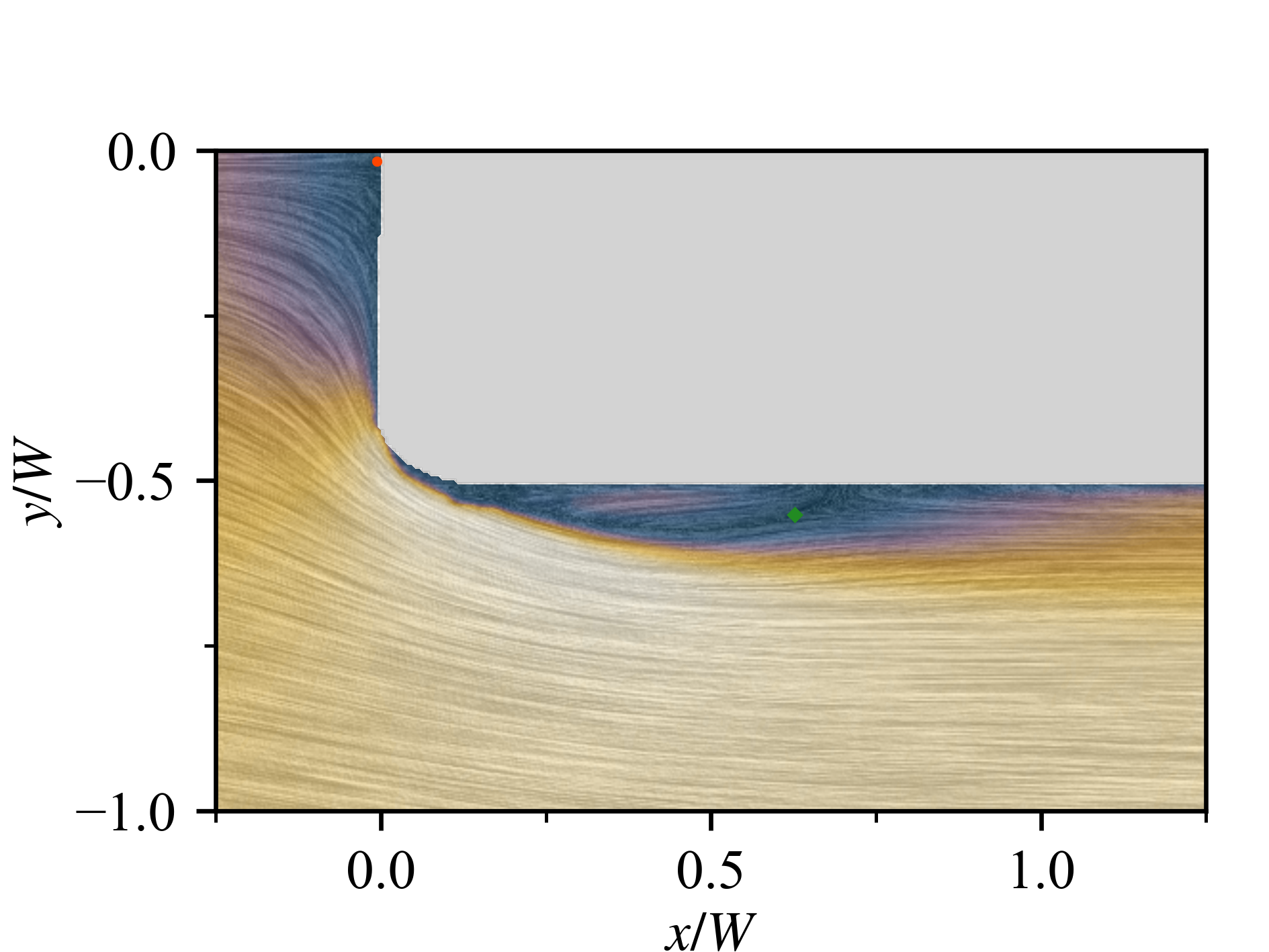}\label{fig:lic_avg_0deg_ACT}}\\
    \subfloat[]{\includegraphics[trim=0cm 0.7cm 0.1cm 0.1cm, clip]{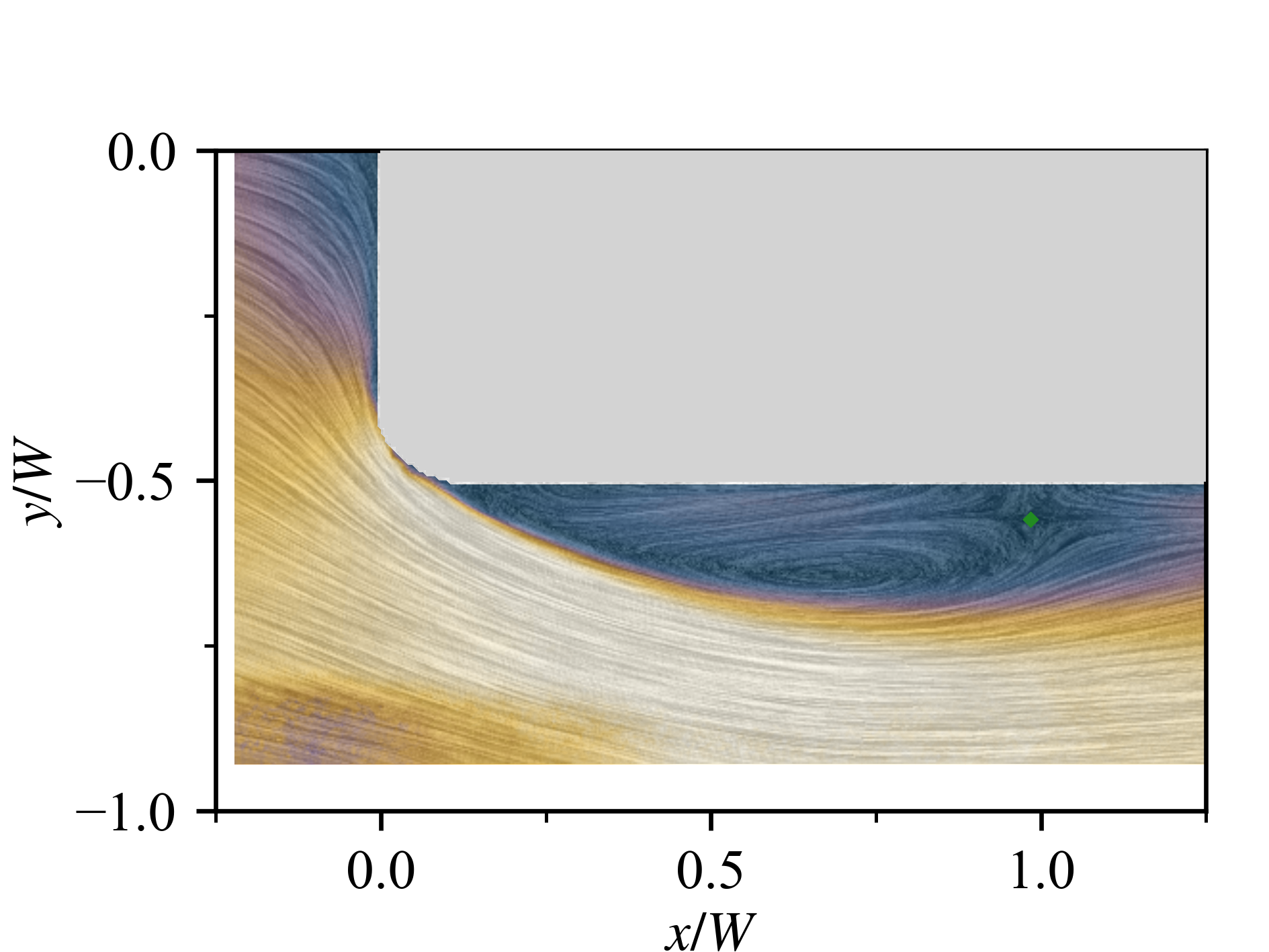}\label{fig:lic_avg_5degL_NA}}\hfill%
    \subfloat[]{\includegraphics[trim=1.3cm 0.7cm 0.1cm 0.1cm, clip]{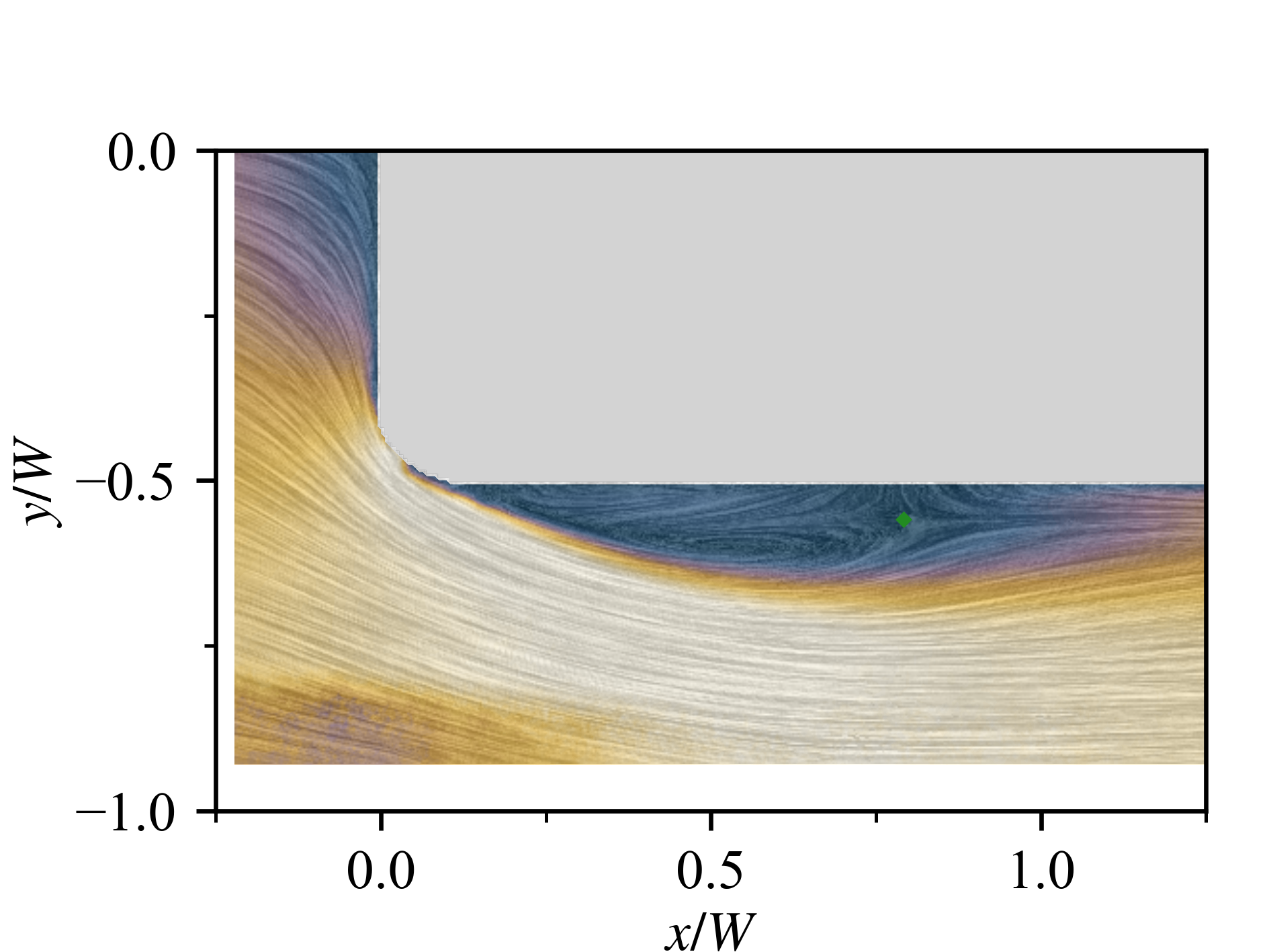}\label{fig:lic_avg_5degL_ACT}}\\
    \subfloat[]{\includegraphics[trim=0cm 0cm 0.1cm 0.1cm, clip]{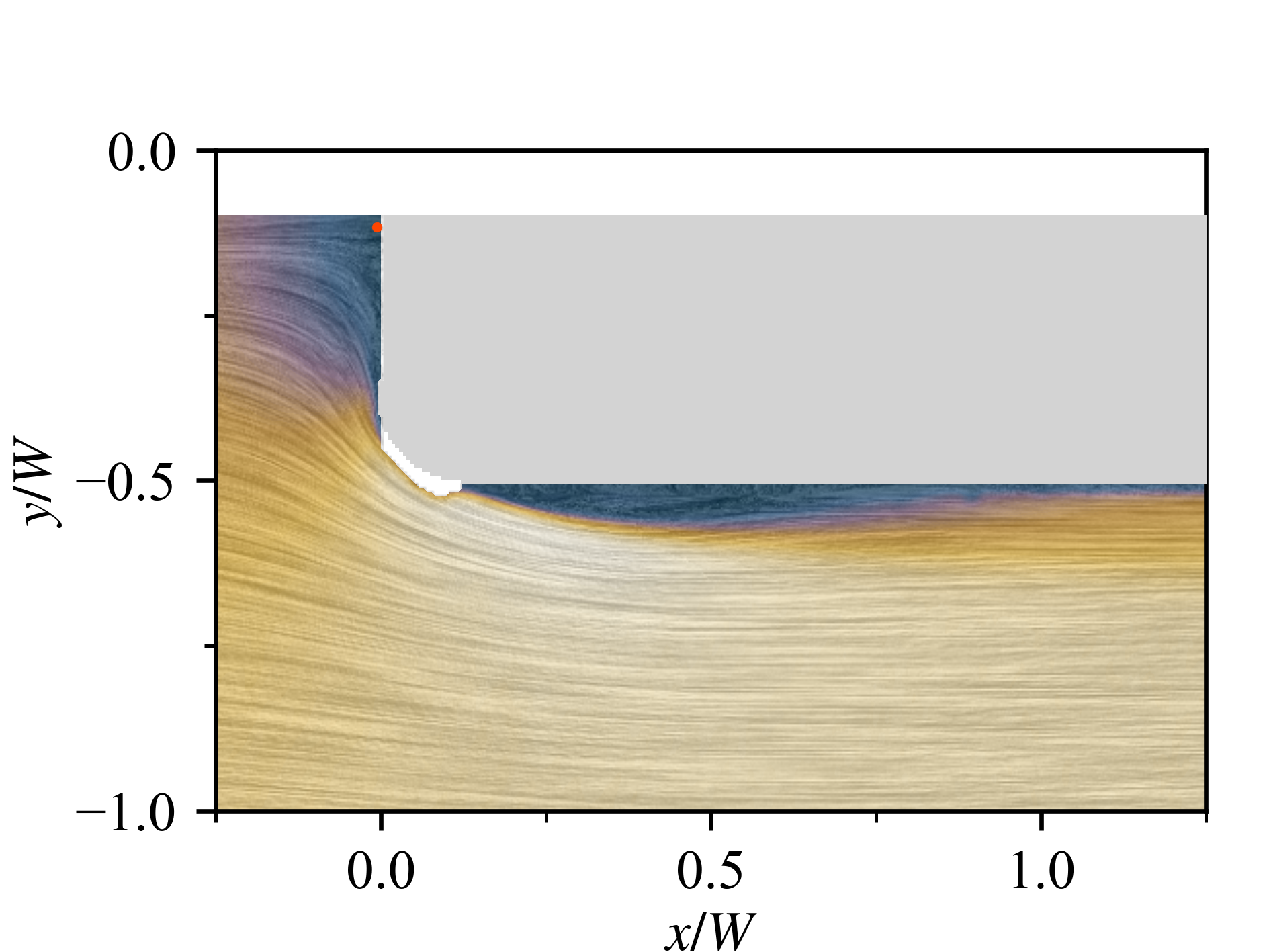}\label{fig:lic_avg_5degR_NA}}\hfill%
    \subfloat[]{\includegraphics[trim=1.3cm 0cm 0.1cm 0.1cm, clip]{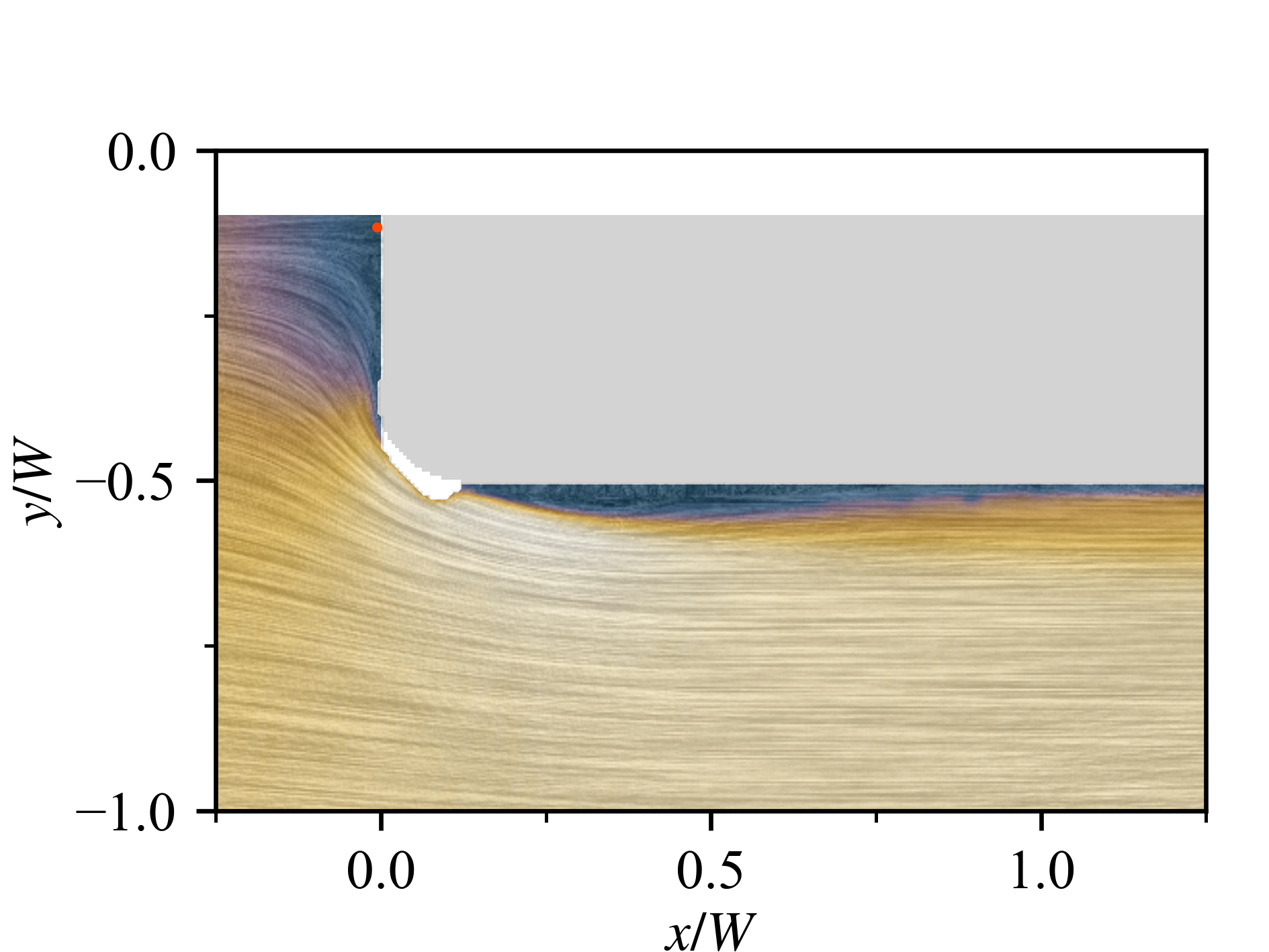}\label{fig:lic_avg_5degR_ACT}}\\
\end{minipage}%
\caption{Visualization of the separation bubble on the \ac{GTS} left side, based on \ac{PIV} measurements. First, second, and third rows show $\alpha=0\degree$, $\alpha=5\degree$, and $\alpha=-5\degree$, respectively. Figures on the left half show the velocity field without actuation, and on the right with actuation. Flow is from left to right. The fields are averaged over 1500 samples. The color represents the normalized velocity magnitude, while the flow direction is rendered using a LIC visualization. The red dots indicate the stagnation points and the green diamonds the rear saddle points. The spatial coordinates are given as fractions of the truck width $W$.}
\label{fig:lic_avg}
\end{figure*}

Figure \ref{fig:lic_avg} presents time-averaged flow fields around the A-pillar for $\alpha= 0\degree$, $5\degree$, and $-5\degree$. The color is representative of the average velocity magnitude, and a line integral convolution \citep[LIC,][]{cabral_imaging_1993} of a white-noise texture along the velocity field shows the flow direction. This visualization technique provides a dense, continuous representation of flow direction and is less sensitive to seeding and clutter than streamline-based visualizations, which quickly become overloaded at high densities. Figures \ref{fig:lic_avg_0deg_NA} and \ref{fig:lic_avg_0deg_ACT} show the velocity fields around the left A-pillar for yaw $0\degree$, averaged over 1500 instantaneous fields. On both plots, the incoming flow slows down and is strongly deviated towards the A-pillar as it approaches the truck front wall. The stagnation point appears at the top of the figures as indicated by the red dot.The flow is then strongly accelerated as it negotiates the rounded corner, eventually leading to flow separation right at the end of the A-pillar.


The recirculation region is delimited from the outer flow by a thin, high-shear layer from the separation point at $(x,y)=(0.1, -0.5)$ until the reattachment at $(x, y)\approx(1.1, -0.5)$. The recirculation region is composed by a main focus rotating counter-clockwise. Comparing the recirculating region on the bottom figure (with actuation) with the top one (baseline case), we clearly see that the plasma actuator has the effect of shrinking the recirculation bubble. Figures \ref{fig:lic_avg_5degL_NA} and \ref{fig:lic_avg_5degR_NA} show the same flow for yaw $5\degree$ and $-5\degree$, respectively. One can readily observe the difference in the dimension of the separation bubble. The actuated case shown in \ref{fig:lic_avg_5degL_ACT} corresponds to a leeward actuation, while \ref{fig:lic_avg_5degR_ACT} describes a windward actuation. Where the bubble is largest (Figure \ref{fig:lic_avg_5degL_NA}), the saddle point delimiting the different recirculation regions is visible and is marked by a green diamond. A secondary recirculation zone appears very close to the truck wall, from the leading edge of the bubble until the saddle point. It rotates clockwise, while the main recirculation area, located a bit downstream and further away from the wall, rotates in a counter-clockwise direction. These features are consistent with those reported by \cite{rao_investigation_2019}. The general structure of the separation bubble is preserved through the actuation, but the saddle point moves upstream from $x\approx1.0$ in the unactuated case to $x\approx0.8$ with actuation. The same phenomenon, although less pronounced, happens at $\alpha=0\degree$. At $\alpha=-5\degree$, the size of the bubble is too small to identify clearly the saddle point with the available resolution.

One can also see how the stagnation point at the truck's front (marked by a red circle) shifts with the yaw angle. For $\alpha=5\degree$ (Figures \ref{fig:lic_avg_5degL_NA} and \ref{fig:lic_avg_5degL_ACT}) the stagnation point moves outside of the displayed area, to the right half of the truck's front ($y/W>0$). Conversely, for $\alpha=-5\degree$ the stagnation point shifts to the left half, around $y/W\approx-0.12$ (Figures \ref{fig:lic_avg_5degR_NA} and \ref{fig:lic_avg_5degR_ACT}). A region around the A-pillar on Figures \ref{fig:lic_avg_5degR_NA} and \ref{fig:lic_avg_5degR_ACT} has been blanked to remove invalid vectors due to the laser reflection caused by the presence of the plasma actuator, making the velocity measurements unusable there. Background effects in \ac{PIV} snapshots have been minimized with the POD-based background removal explained in Section \ref{sec:methodology}, but persist in that case due to the model inclination, which increases light scattering.

\begin{figure*}[t]
\centering
\begin{minipage}{15cm}
    \centering
     \subfloat[]{%
      \includegraphics[trim=0cm 0cm 0.2cm 0.6cm, clip]{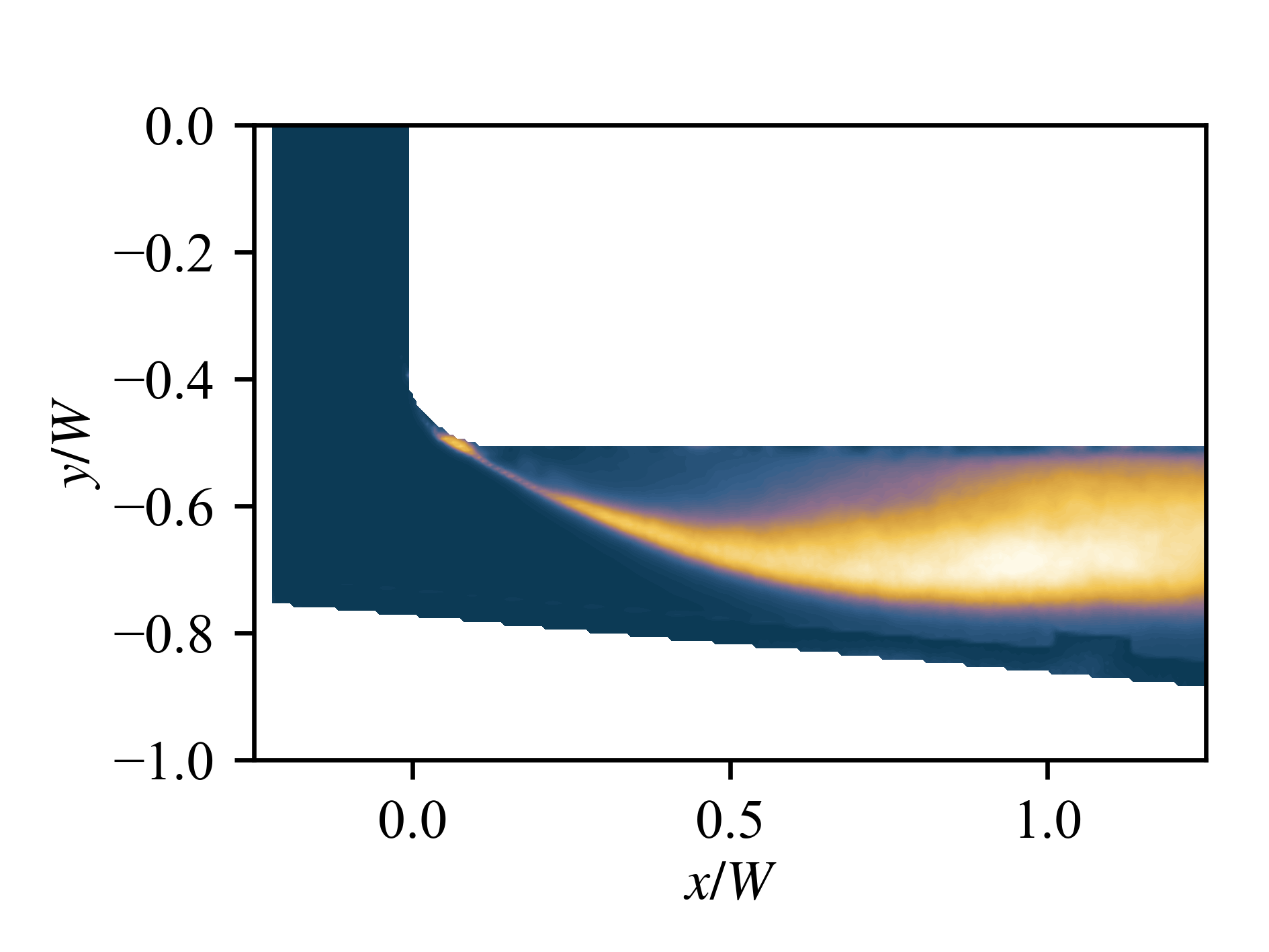}%
        \label{fig:tke_5degL_na}%
    }\hfill
    \subfloat[]{%
        \includegraphics[trim=1.4cm 0cm 0.2cm 0.6cm, clip]{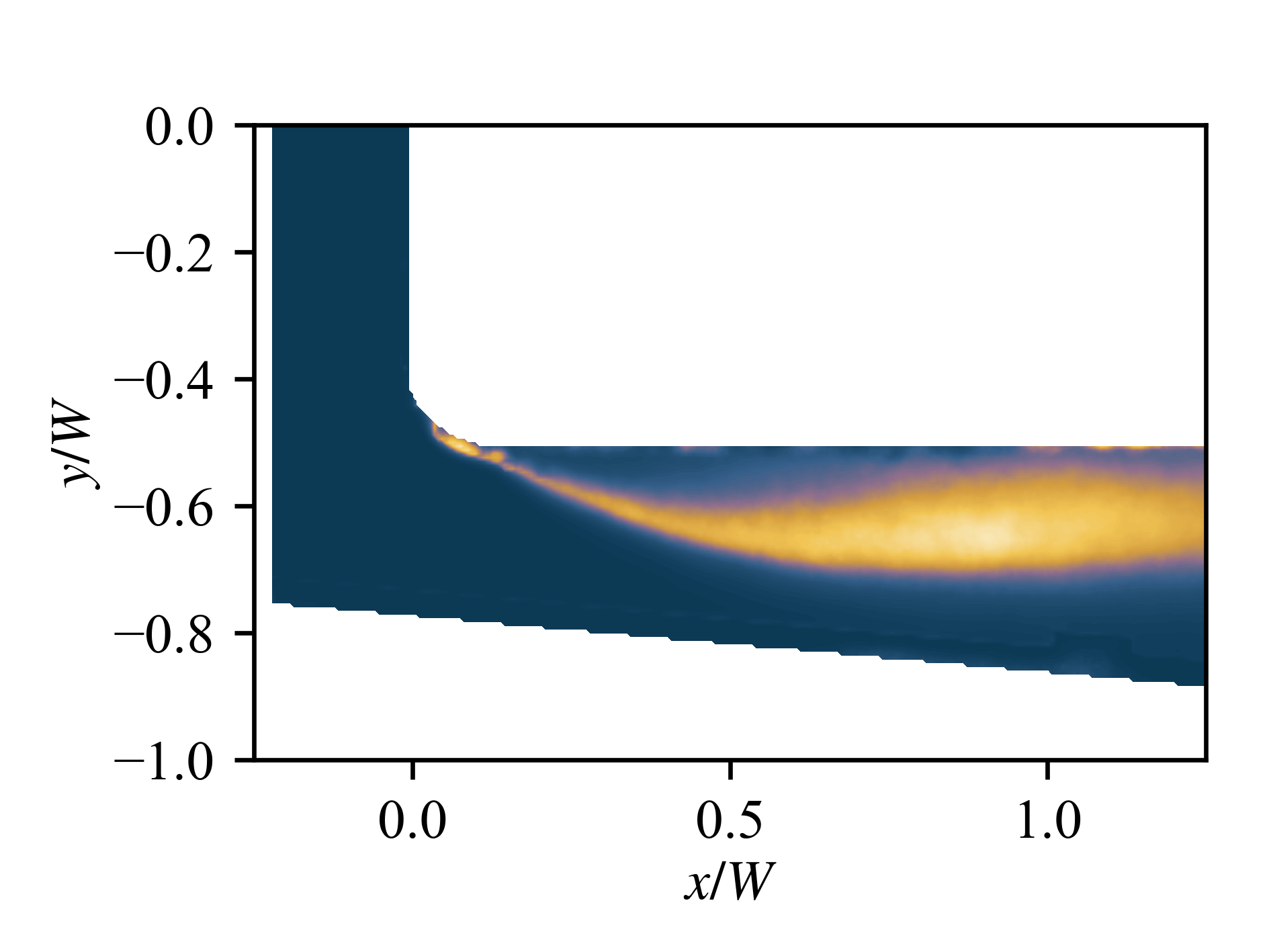}%
        \label{fig:tke_5degL_act}%
    }\hfill
\end{minipage}%
\begin{minipage}{3cm}
    \centering
    \includegraphics{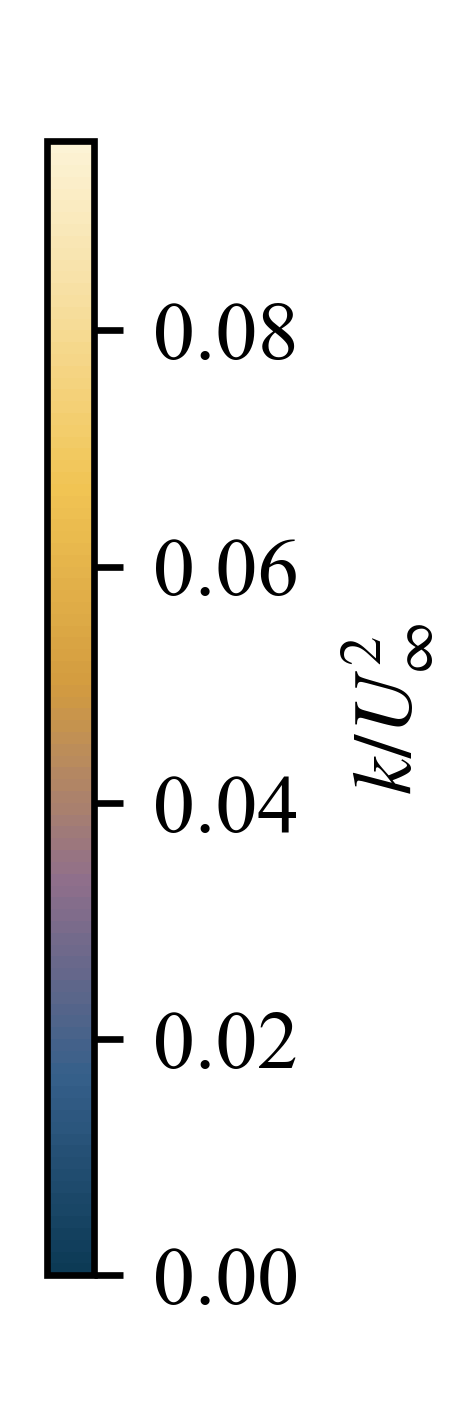}
\end{minipage}
\caption{In-plane turbulent kinetic energy $k=\frac{1}{2}(\overline{u'^2}+\overline{v'^2})$ for $\alpha=5 \degree$: (a) case without actuation (b) case with actuation.}
\label{fig:tke}
\end{figure*}

The effect of plasma actuation on higher-order statistics can be exemplified by a comparison of the turbulent kinetic energy field $k$ with and without actuation at a given yaw angle. In Figure \ref{fig:tke}, we report the normalized in-plane turbulent kinetic energy measured at $\alpha=5\degree$. Not only does the actuation reduces de size of the separation after the A-pillar, it also decreases the amount of turbulent kinetic energy, thus reducing the extraction of momentum from the mean flow towards the turbulent cascade, further explaining the drag reduction \citep{bonnavion_use_2022}.

\begin{figure}[h]
    \centering
    \includegraphics{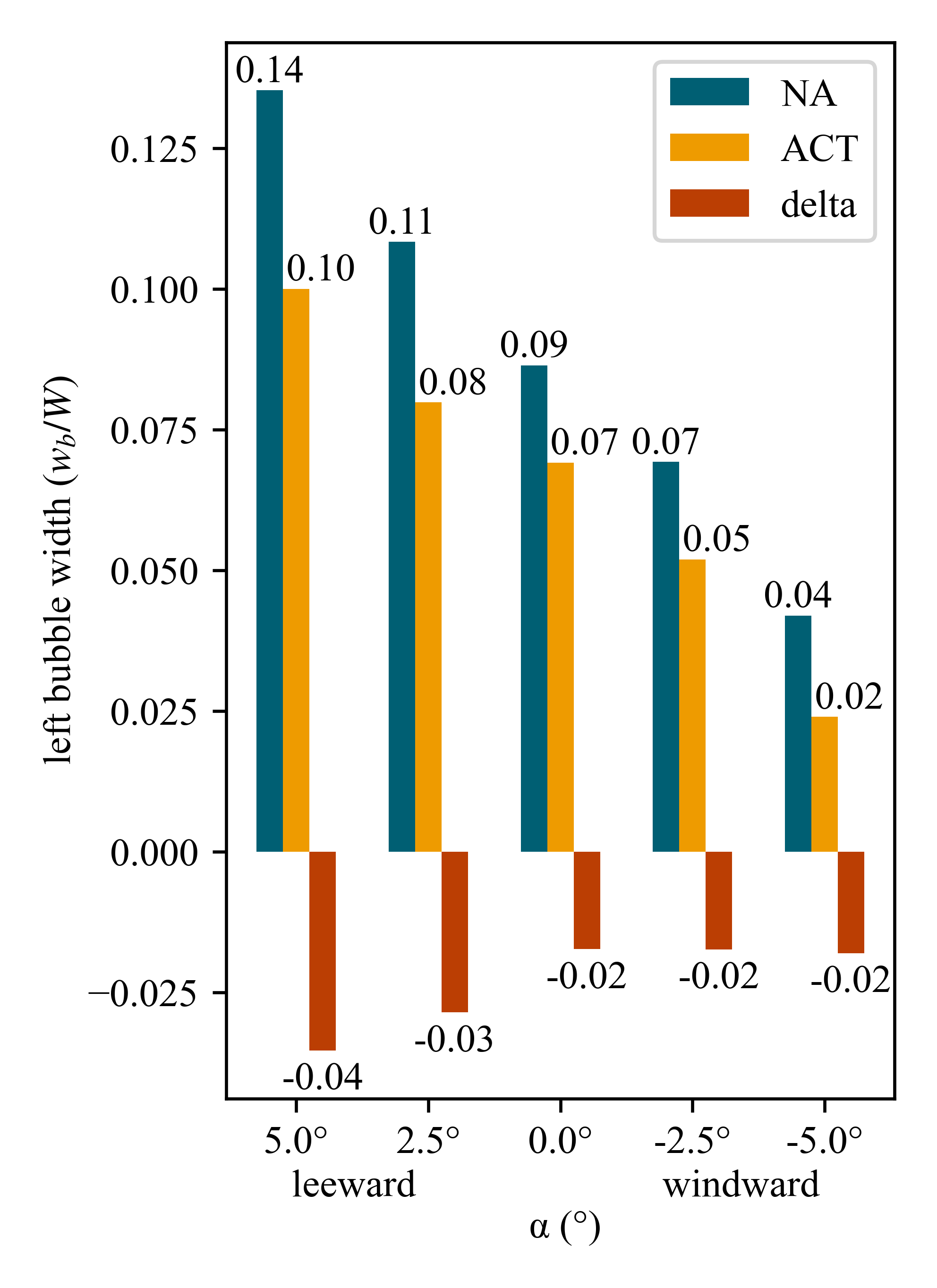}
    \caption{Bubble width as a function of the yaw angle, non-dimensionalized by the truck width.}
    \label{fig:bubble_width}
\end{figure}

\begin{figure}[h]
    \centering
    \includegraphics{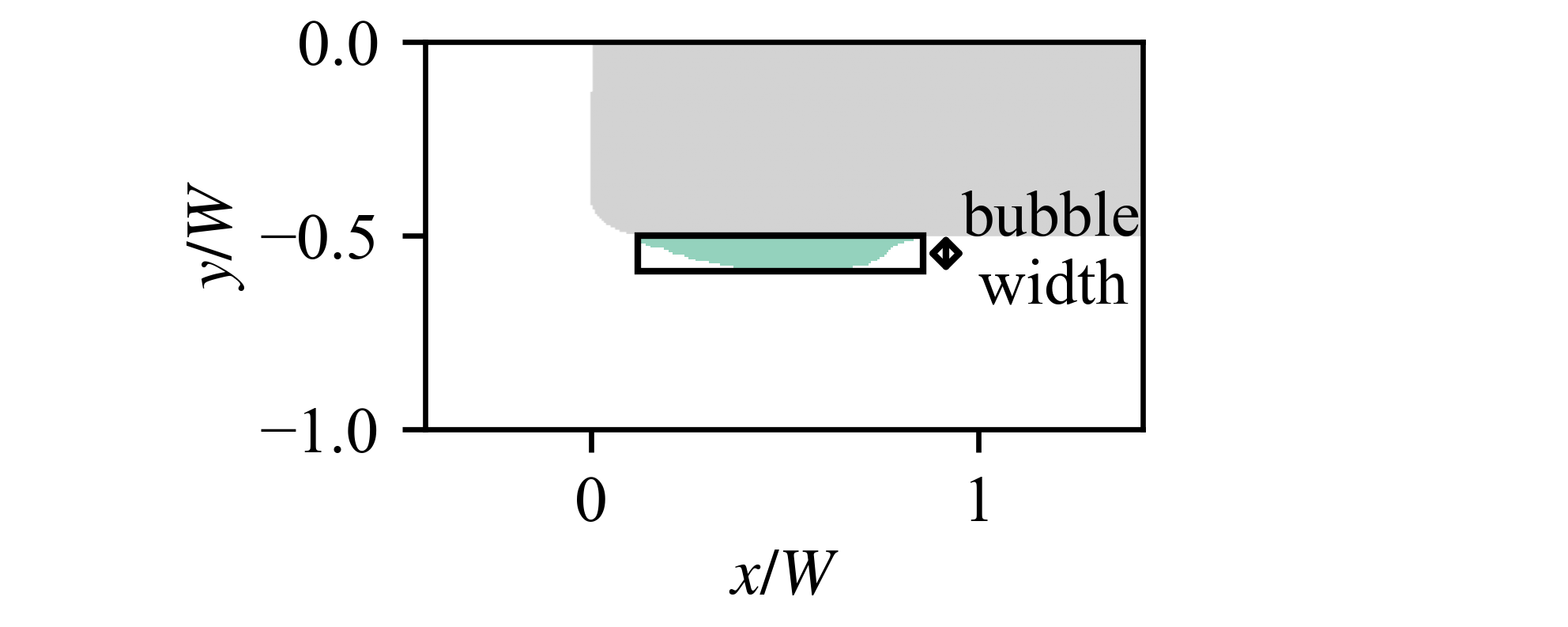}
    \caption{The truck side with the area of at least 50\% backflow highlighted in turquoise and the minimal circumscribing rectangle defining the bubble width in black.}
    \label{fig:bubble_labeled_250313_0deg_IW32}
\end{figure}

Figure \ref{fig:bubble_width} allows for a comparison of the bubble width among different yaw angles with or without actuation. The bubble width and length are calculated as the dimensions of the smallest rectangle aligned with the wall circumscribing the area displaying backflow in at least 50\% of the valid snapshots, as illustrated on Figure \ref{fig:bubble_labeled_250313_0deg_IW32}.
Backflow is defined whenever the streamwise velocity component is smaller than a threshold, i.e. $u < \varepsilon$, where $\varepsilon=0.1~\text{m/s}$ accounts for measurement uncertainty in \ac{PIV} data. If several disjunct backflow areas are identified, the largest one is considered. It is important to note that the dimensions of the bubble are rather small: the smallest bubble is $1.7~\unit{mm}$ wide. We are thus considering an optical measurement very close to the wall.

The increase in the bubble width due to yaw is practically linear in the baseline case. In the actuated case, the deviation from a linear fit is larger; the value of the least squares residual after a linear regression is about 3 times that for the baseline case. When looking at the difference between both cases represented by the lower bars in Figure \ref{fig:bubble_width}, the bubble reduction augments linearly with the yaw angle on the leeward side, but remains approximately constant on the windward side. This result is consistent with the force measurements:
\begin{itemize}
    \item Larger separation bubbles induce more axial force, as the separation contributes to the apparent width of the truck, i.e., the flow has to go around the \ac{GTS} body and around the separation bubble, as if the truck were wider. This is the case when yaw increases.
    \item Leeward actuation is more effective than windward actuation in thinning the separation bubble, as it is in reducing the axial force.
    \item As expected from potential flow theory \cite[see, e.g.][]{kundu_ideal_2012}, the vorticity of the recirculation bubble is responsible for a suction force acting on the corresponding wall. Actuating on the windward recirculation bubble increases the difference between the two suctions on the leeward and windward side, thus increasing the side force.
\end{itemize}

\begin{figure}[h]
    \centering
    \includegraphics{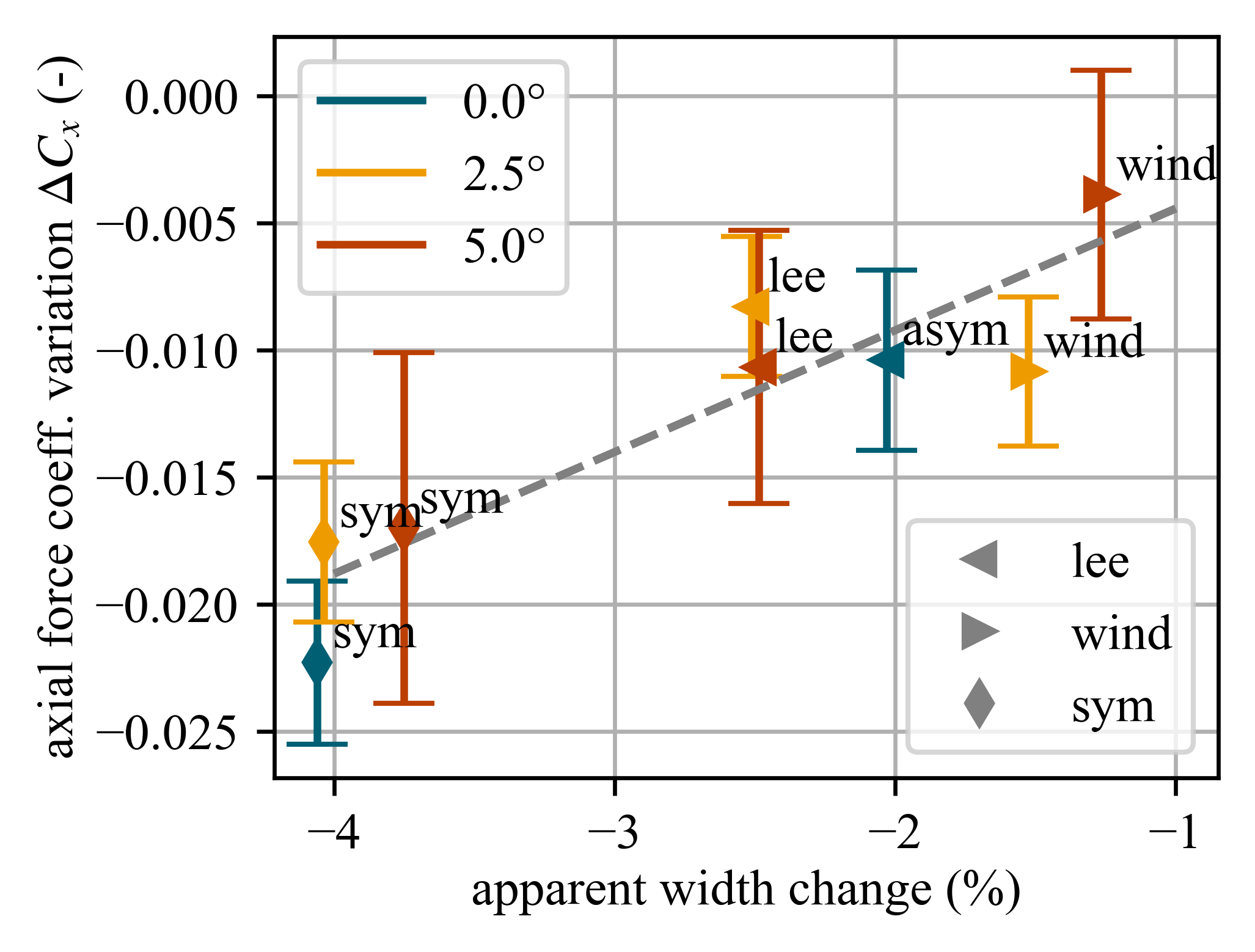}
    \caption{Change in axial force coefficient as a function of the relative change in apparent width in the spanwise axis. A linear fit for all points is shown in gray.}
    \label{fig:width_relchange_axial_var}
\end{figure}

 A relation between the bubble size and the axial force can be sketched by projecting the change of bubble width in the spanwise plane. In this case, we need to sum the widths of the leeward and the windward bubbles for a given yaw angle and project them on the spanwise axis. The result is shown in Figure \ref{fig:width_relchange_axial_var}. The variation of the spanwise component of the apparent width correlates with the axial force reduction. A straight fit is indicated for illustrative purposes, but does not indicate a linear relationship, as the number of datapoints as well as the coefficient of determination are really low ($R^2=0.82$). However, one can see that the symmetrically-actuated cases are at the bottom left, as they display the largest force reduction and width change, and the least effective actuation case (windward at $5\degree$ yaw) is in the upper right corner.

\begin{figure*}[!h]
    \centering
    \includegraphics{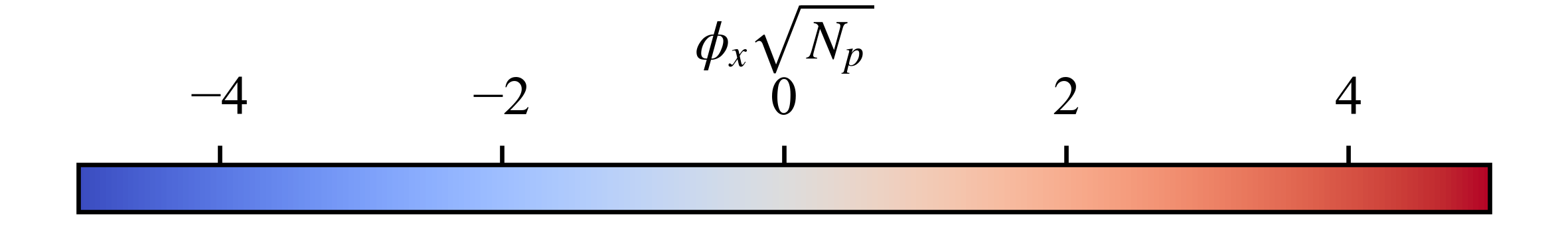}\\
    \subfloat[]{\includegraphics{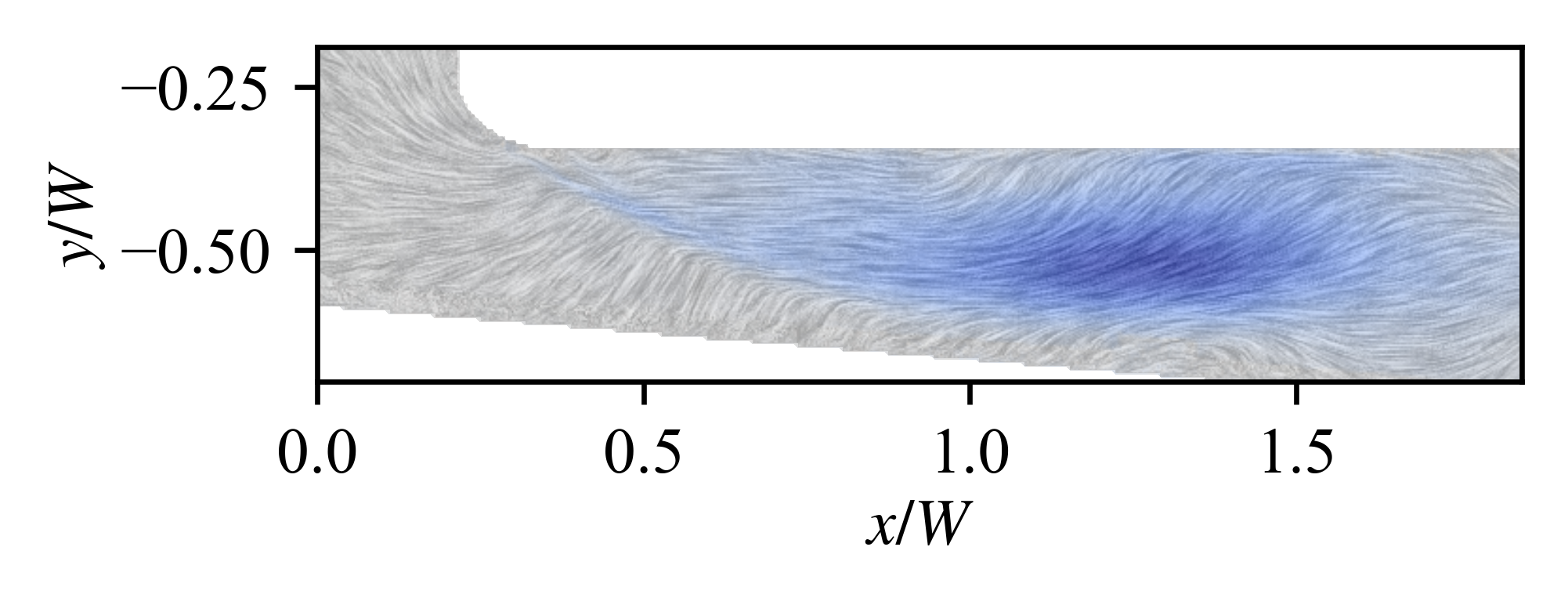}\label{fig:PODrecon100_5degL_IW32_NA_mode_1}}%
    \subfloat[]{\includegraphics{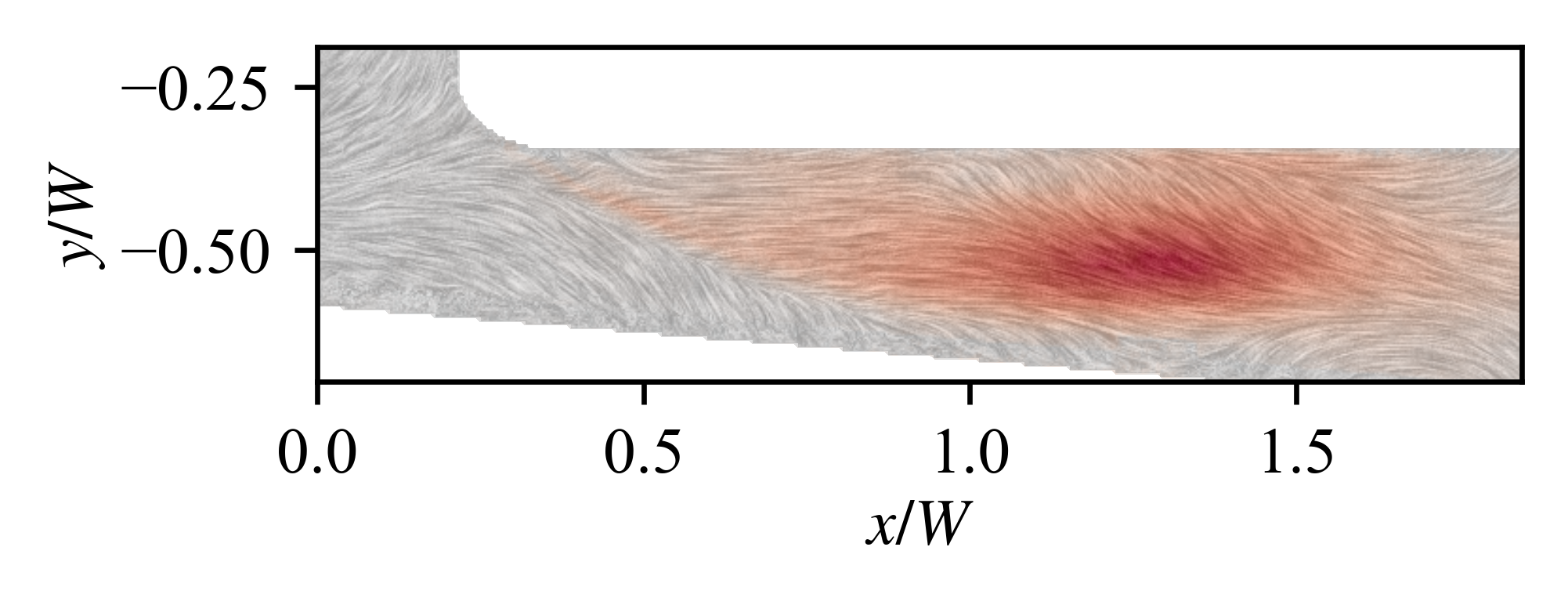}\label{fig:PODrecon100_5degL_IW32_ACT_mode_1}}\\
    \subfloat[]{\includegraphics{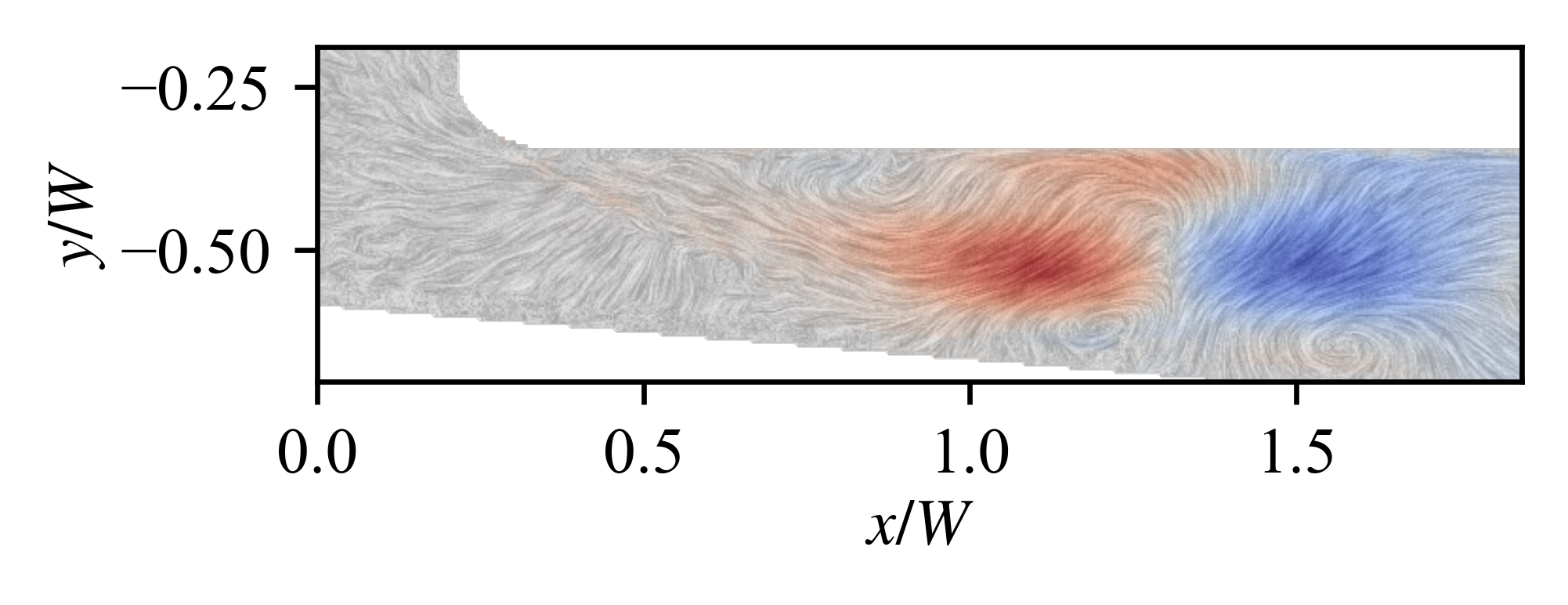}\label{fig:PODrecon100_5degL_IW32_NA_mode_2}}%
    \subfloat[]{\includegraphics{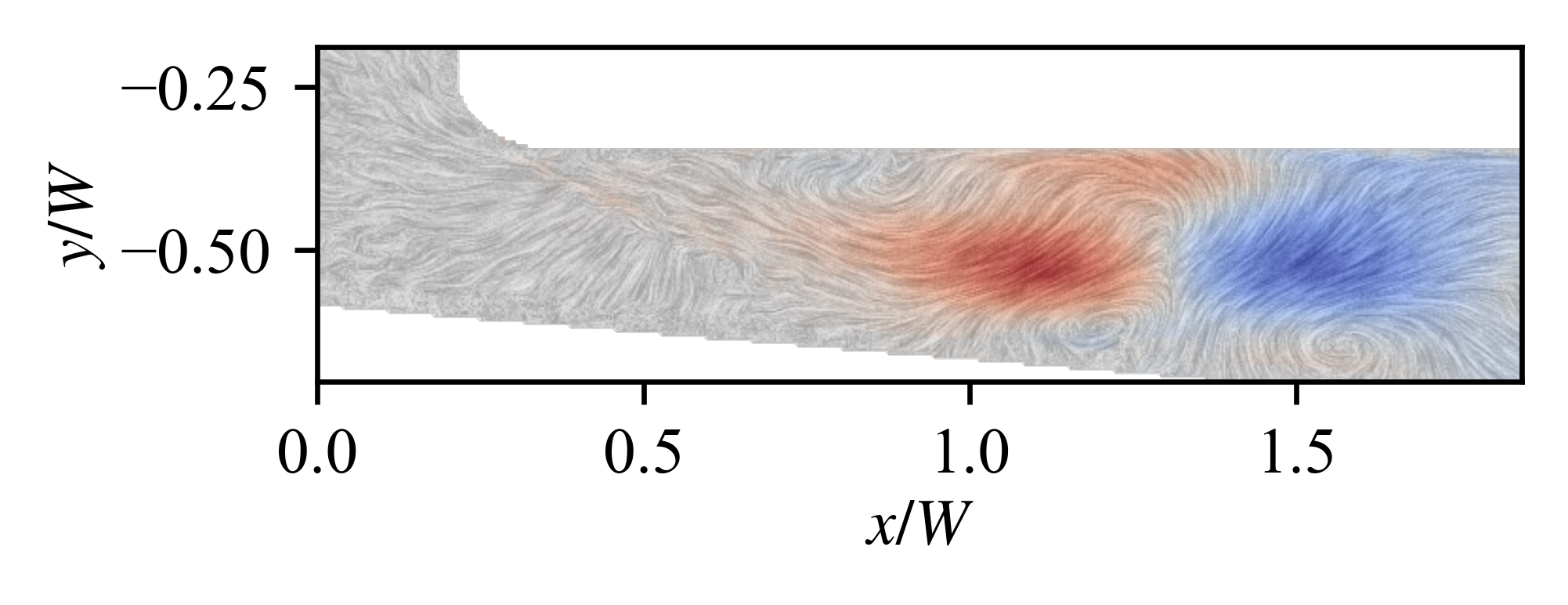}\label{fig:PODrecon100_5degL_IW32_ACT_mode_2}}\\
    \subfloat[]{\includegraphics{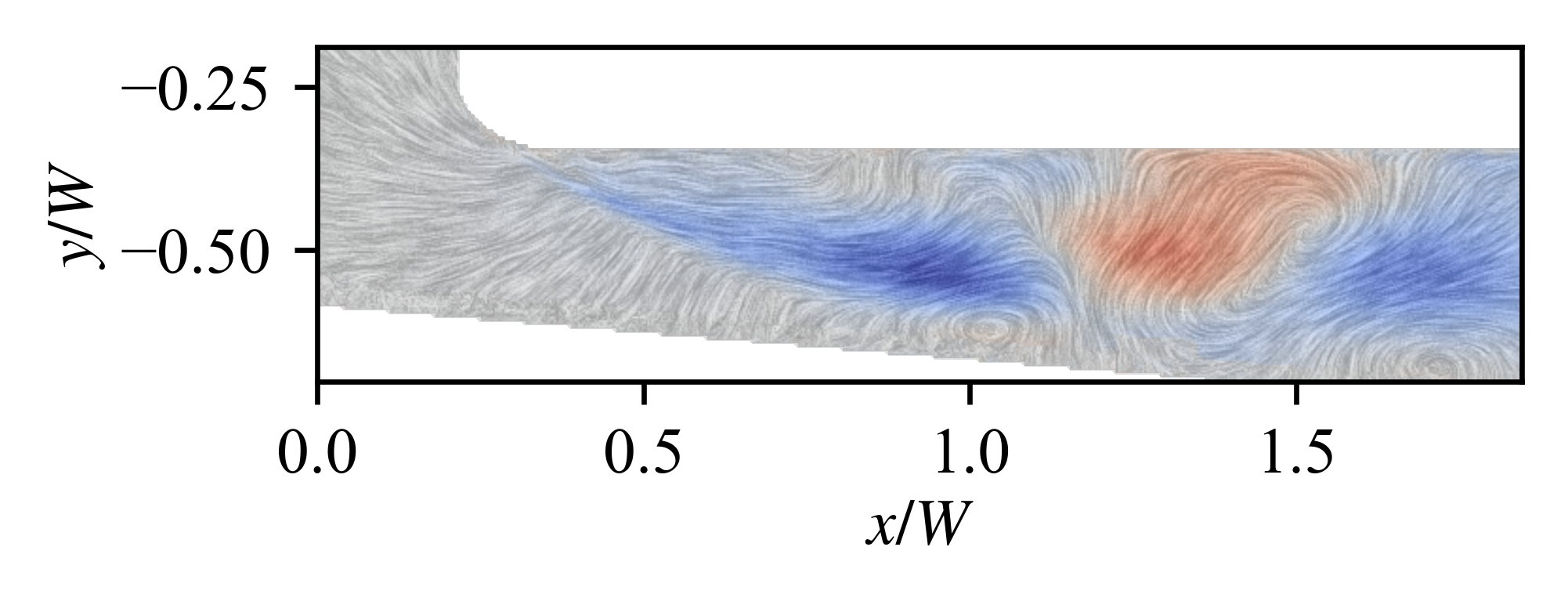}\label{fig:PODrecon100_5degL_IW32_NA_mode_3}}%
    \subfloat[]{\includegraphics{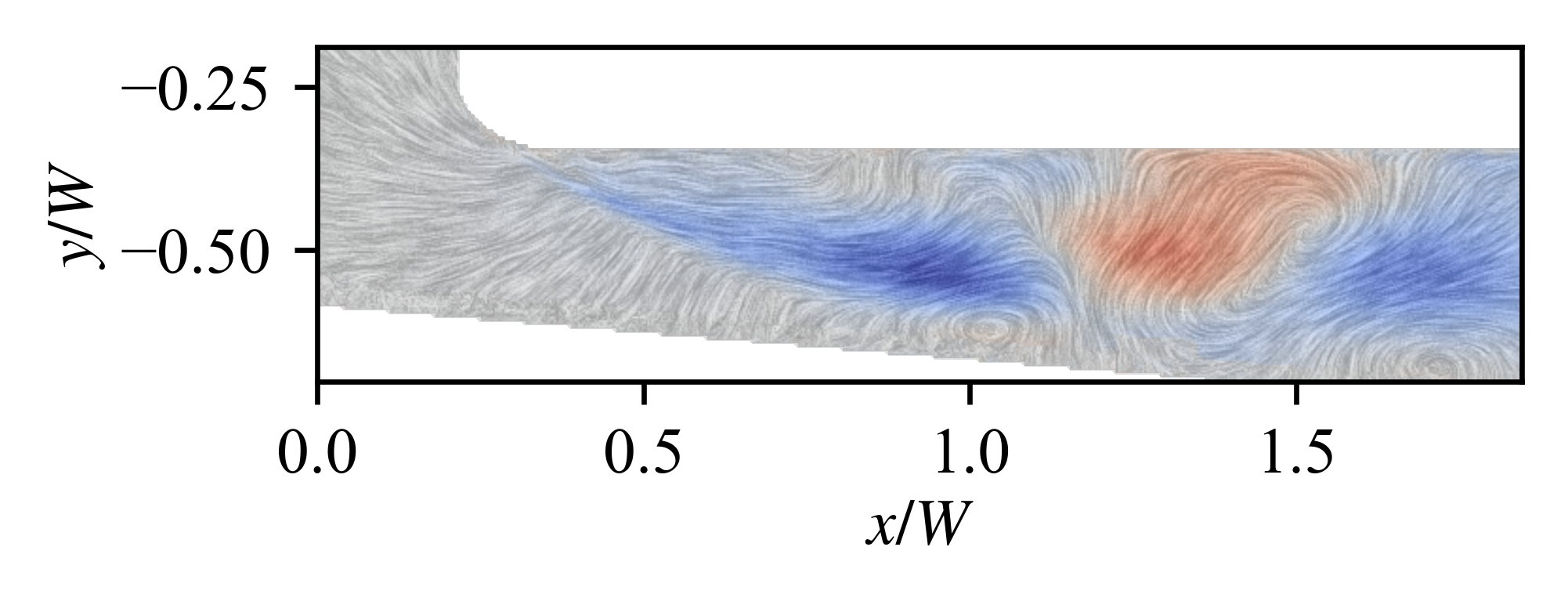}\label{fig:PODrecon100_5degL_IW32_ACT_mode_3}}\\
    \caption{POD modes at yaw $5\degree$. Figures \ref{fig:PODrecon100_5degL_IW32_NA_mode_1}, \ref{fig:PODrecon100_5degL_IW32_NA_mode_2} and \ref{fig:PODrecon100_5degL_IW32_NA_mode_3} show the three first modes without actuation and Figures \ref{fig:PODrecon100_5degL_IW32_ACT_mode_1}, \ref{fig:PODrecon100_5degL_IW32_ACT_mode_2} and \ref{fig:PODrecon100_5degL_IW32_ACT_mode_3} with the plasma actuator on. Color shows the streamwise component of the spatial modes $\phi_x$.}
    \label{fig:PODrecon100_5degL_IW32}
\end{figure*}

After analyzing the average flow features, it is worth inspecting the statistical flow organization and the kinematics of the separation bubble. To this purpose, we analyze \ac{PIV} snapshots with POD \citep{berkooz_proper_1993} of the velocity fluctuations. Figure \ref{fig:PODrecon100_5degL_IW32} reports the first three spatial modes obtained from the POD for the non-actuated and the leeward-actuated case for a yaw angle of $5\degree$. The color maps the streamwise component of each spatial mode $\phi_x$. Since the POD modes obtained from the snapshot matrix have unitary norm, they are multiplied by the square root of the number of grid points $N_p$ to obtain a representation with values of order 1. A LIC is again employed to visualize the streamlines of the velocity fluctuations over the domain. One can see that the modes in both cases are similar. The first mode shows the extension/retraction of the bubble, like a pumping mechanism, while higher-order modes present higher-frequency sheddings on the edge of the recirculation bubble. While the actuation changes the width of the bubble, moving the features closer to the wall, the similarity between the non-actuated and the actuated modes suggests that the kinematics of the bubble is not affected by the actuation.

Previous literature on the control of laminar separation bubbles with plasma actuators \citep{yarusevych_effect_2017} indicates that the bubble size, the coherent structures in the separated shear layer, and the stability characteristics depend strongly on actuation frequency and amplitude. The present finding about POD modes not being affected by the actuation suggests that pulsed or time-modulated plasma actuation could allow controlling the bubble dynamics, possibly leading to improved control capabilities.

\section{Conclusion}

Planar \ac{PIV} confirms the presence of a lateral separation bubble on both sides of the \ac{GTS} cabin and show that linear \ac{DBD} plasma actuators placed on the A-pillars effectively reduce the bubble size within the tested yaw range $[-5\degree, 5\degree]$. Force measurements demonstrate that up to $\alpha \approx 8\degree$ the axial force can be effectively reduced by actuating on each side independently or simultaneously. The strong impact of windward actuation on drag in light crosswind conditions constitutes a key novelty with respect to previous experimental studies, which focused on leeward actuation. Beyond this angle, the windward actuation becomes ineffective, and leeward actuation alone yields a drag reduction equivalent to symmetric actuation.

The lateral force is also influenced by the action of the plasma. In crosswind conditions, windward actuation tends to increase the lateral force magnitude, whereas leeward actuation reduces it. When combined, the effect of the windward actuation dominates, resulting in a net increase in the lateral force, although not as sharp as with the windward actuator alone.
The limited magnitude of the variations experienced by the lateral force due to the plasma actuation reflects the relatively small size of the A-pillar separation region with respect to the truck side area. However, this effect could become stronger with improved actuation authority or considering other geometries.

Finally, a control strategy adapting the actuation to the yaw angle is proposed. It consists of turning off the windward actuator when the yaw angle exceeds a specific threshold, thereby optimizing aerodynamic performance and stability as well as control power expenses.

The findings of this research allow for a better understanding of the physics of the actuation
mechanism and its effect on the flow separation. We
acknowledge the long road towards higher \ac{TRL} demonstrations leading to the application of plasma actuators to real heavy-duty vehicles. Further work should focus on exploring the control signal parameter space to maximize control authority and energy efficiency. In the field of aeronautics, previous literature already demonstrated the possibility of implementing plasma actuators in a higher \ac{TRL} environment \citep{su_uav_2018, sekimoto_-flight_2022} supporting the hypothesis that \ac{DBD} plasma actuators can also be leveraged for real-world truck applications.


%
%

%

\begin{acknowledgments}
This activity is part of the project ACCREDITATION (Grant No TED2021-131453BI00), funded by MICIU/AEI/10.13039/501100011033 and by the “European Union NextGenerationEU/PRTR”.
\end{acknowledgments}

\section*{Author declarations}

\subsection*{Conflict of Interest}
The authors have no conflicts to disclose.

\subsection*{Author contributions}
\textbf{Lucas Schneeberger}: Methodology, Software, Validation, Formal Analysis, Investigation, Data Curation, Writing/Original Draft Preparation, Visualization.
\textbf{Stefano Discetti}: Conceptualization, Methodology, Resources, Writing/Review \& Editing, Supervision, Project Administration, Funding Acquisition.
\textbf{Andrea Ianiro}: Conceptualization, Methodology, Resources, Writing/Review \& Editing, Supervision, Project Administration, Funding Acquisition.

\section*{Data availability}
The data that support the findings of this study are available from the corresponding author upon reasonable request.


\bibliography{PhD}

\end{document}